\documentclass[12pt,letterpaper]{article}

\usepackage{amsmath}
\usepackage{amsfonts}
\usepackage{amssymb}
\usepackage{amsthm}
\usepackage{graphicx}
\usepackage{mathrsfs}
\usepackage{xcolor}
\usepackage{url}
\usepackage{todonotes}
\usepackage{verbatim}
\usepackage{titletoc}
\usepackage{color}
\usepackage{pdflscape}

\usepackage[letterpaper]{geometry}
\usepackage{setspace}
\allowdisplaybreaks

\usepackage[authoryear]{natbib}

\usepackage[subtle]{savetrees}
\usepackage[small,compact]{titlesec}

\newtheoremstyle{break}
  {\topsep}{\topsep}%
  {\itshape}{}%
  {\bfseries}{}%
  {\newline}{}%
\theoremstyle{break}

\newtheorem{proposition}{Proposition}
\newtheorem{assumption}{Assumption}
\newtheorem{lemma}{Lemma}

\usepackage[font=small, labelfont=bf]{caption}

\DeclareMathOperator{\argmin}{argmin}

\begin{document}

\title{Hybrid Confidence Intervals for Informative Uniform Asymptotic Inference After Model Selection\thanks{
I thank Arun Kuchibhotla (Carnegie Mellon University) for very helpful comments and discussions.}} 


\author{Adam McCloskey\footnote{Department of Economics, University of Colorado, adam.mccloskey@colorado.edu}
}

\maketitle

\begin{abstract}
I propose a new type of confidence interval for correct asymptotic inference after using data to select a model of interest without assuming any model is correctly specified.  This hybrid confidence interval is constructed by combining techniques from the selective inference and post-selection inference literatures to yield a short confidence interval across a wide range of data realizations.  I show that hybrid confidence intervals have correct asymptotic coverage, uniformly over a large class of probability distributions that do not bound scaled model parameters.  I illustrate the use of these confidence intervals in the problem of inference after using the LASSO objective function to select a regression model of interest and provide evidence of their desirable length and coverage properties in small samples via a set of Monte Carlo experiments that entail a variety of different data distributions as well as an empirical application to the predictors of diabetes disease progression.

\bigskip

\noindent \textsc{Keywords:~Confidence Interval, Selective Inference, Post-Selection Inference, LASSO, Uniform Asymptotics, Misspecification}
\end{abstract}

\thispagestyle{empty}
\setcounter{page}{0}

\newpage

\section{Introduction}\label{sec:intro}

A large portion of the statistics literature in recent years has been dedicated to advancing inference methods that are valid after using data to select a model of interest without assuming the correct specification of any model in the selection set.  Many, if not most, of these methods can be roughly broken down into two strands:~methods that are valid for a selected parameter conditional on the model that is selected using a particular model selection criterion and methods that are unconditionally valid irrespective of the particular model selection criterion used.  The former is often referred to as ``selective inference'' (e.g.,~\citealp{LSST16}) and the latter is often referred to as ``post-selection inference'' (PoSI) (e.g.,~\citealp{BBBZZ}), terms I will use throughout this paper.  

Apart from the differences in coverage guarantees from confidence intervals (CIs) constructed from these two approaches, they also feature complementary strengths and weaknesses in terms of informativeness.  While selective CIs tend to be short when the model selected by the data is selected with (unconditional) high probability, they can become exceedingly wide when this selection event occurs with low probability.  Indeed, standard selective CIs approach naive CIs (with incorrect coverage for the selected parameter) based upon inverting $t$-tests while ignoring data-driven model selection when the selection event occurs with high probability (\citealp{AKM20}).  On the other hand, their expected length may be infinite (\citealp{KL20}).  PoSI CIs do not suffer this latter drawback but can be very conservative in scenarios where the model selection criterion is known and the model selected by the data is selected with high probability, leading to coverage probabilities well in excess of their nominal levels and unnecessarily wide CIs.

In this paper, I propose a new class of hybrid (HySI, short for ``hybrid selective inference'') CIs for inference after model selection that aims to draw on the complementary informativeness strengths of selective and PoSI CIs.  Motivated by the fact that selective CIs can become extremely wide when the model selected by the data is selected with low (unconditional) probability, by relaxing the conditional coverage requirement it may be possible to attain CIs with guaranteed unconditional coverage and better length properties.  Although PoSI CIs have guaranteed unconditional coverage, selective CIs are substantially shorter when the model selected by the data is selected with high probability.  In order to use the relative strengths of these two CIs in terms of their lengths, the HySI approach to CI construction uses the data to transition a PoSI CI toward a selective CI when the latter is short.  The HySI CIs introduced here make use of similar reasoning to the hybrid CIs of \cite{AKM20} and \cite{AKM20b}, but applied to a model selection framework that generalizes that of \cite{AKM20} to incorporate many popular model selection criteria used in linear regression.

Like selective CIs, but unlike PoSI CIs, HySI CIs require knowledge of the model selection criterion used by the researcher and the model selected by the data. The HySI approach relaxes the conditional (on the selected model) coverage requirement of standard selective CIs to produce CIs that are unconditionally valid.  This unconditional coverage guarantee is analogous to the unconditional coverage guarantee of PoSI CIs with one difference:~since we know the model selection criterion being used when constructing HySI CIs, we attain validity ``on average'' across the models potentially selected using this particular model selection criterion, rather than across all possible model selection criteria.\footnote{Selective CIs posess the same unconditional coverage properties as HySI CIs.  For brevity, we refer the interested reader to \cite{TTLT16} and \cite{TRTW18} for detailed discussions on the interpretation of unconditional coverage when the model selection criterion is known.}  The relaxation of the conditional coverage requirement can be viewed as an alternative approach for improving the length properties of selective CIs to standard sample splitting, the ``data-carving'' approach of \cite{FST17} or the randomized response approach of \cite{TT18}.  However, in contrast to these latter approaches, the HySI approach does not discard or add noise to the data in the model selection stage.  It therefore yields valid inference on an object of interest that is selected from the full set of data rather than a fraction of it.

The HySI approach to CI construction improves upon the typical length properties of both the selective and PoSI CIs by modifying the conditioning event of selective CIs to restrict the selected parameter of interest to lie within a PoSI CI with a higher coverage probability.  Since this latter addition to the conditioning event is not necessarily satisfied by the data, the HySI approach takes the selective CI based upon this event but modifies which truncated normal quantiles are used in its constructoin to maintain correct coverage.  By construction, the maximum length of the HySI CI is bounded above by the length of the corresponding higher coverage PoSI CI, yielding finite expected length and breaking the negative result of \cite{KL20}.  At the same time, HySI CIs can be configured to \emph{nearly} approach naive CIs when the model selected by the data is selected with high probability in the sense that they approach naive CIs of a slightly higher coverage level.\footnote{See the discussion at the end of Section \ref{sec:theory} for details.}

Under a strengthening of the general model selection framework of \cite{MXT18} and an assumption implying the existence of a uniformly asymptotically valid PoSI CI, I establish the uniform asymptotic validity of the HySI CIs I propose.  Standard uniform laws of large numbers and central limit theorems (CLTs) and results in e.g.,~\cite{KBB18} and \cite{BPS20} can be used to verify these assumptions in linear regression model selection contexts.  In the absence of knowledge of a ``true'' model, these assumptions typically require the empirically-relevant setting of random, rather than fixed, regressors.  As an illustration, I show how to verify these assumptions in the context of performing inference on a population regression coefficient for a predictor of interest after using the LASSO objective function to select the control variables that enter the regression.  Importantly, this framework does not impose distributional assumptions on the data or that any parameters are known a priori.

The uniform asymptotic validity results can be applied to many other examples such as inference in a linear regression model selected by LASSO with randomized cross-validation, a randomized information criterion, a fixed number of steps along the forward stepwise (FS) and least angle regression (LAR) algorithms and along the solution path of LASSO.  Unlike the point-wise asymptotic results of~e.g.,~\cite{TT17}, the \emph{uniform} asymptotic results I establish in this paper provide better approximations to finite sample coverage across a broad range of data-generating processes (DGPs).  And unlike the results of~e.g.,~\cite{TRTW18} or \cite{AKM20b} but in line with e.g.,~\cite{BPS20}, the uniform asymptotic validity results I establish in this paper do not require one to bound the magnitude of various parameters such as (scaled) population regression coefficients.

Finally, I investigate the finite-sample performance of the HySI CIs relative to the selective CIs of \cite{LSST16}, standard CIs formed after sample splitting and the PoSI CIs of \cite{BPS20} when using LASSO as the model selection criterion in a set of Monte Carlo experiments and an empirical data application.  In the Monte Carlo experiments, I draw data from a variety of distributions in a small sample setting to examine how well the uniform asymptotic results translate to challenging finite-sample settings that significantly depart from Gaussianity.  Under a wide range of values for the LASSO penalty parameter, I find that the length distribution of HySI CIs compares very favorably to those of selective and PoSI CIs.  The length gains relative to selective CIs are acutely pronounced at higher quantiles of the relative length distributions.  As a stark example, the $95^{th}$ quantile of the length distribution of the selective CIs can be more than 34 times larger than that of the HySI CIs under the Monte Carlo designs I examine.  In addition, I find that the HySI CIs exhibit approximately correct finite-sample coverage.  In the empirical application, I analyze the diabetes dataset from \cite{EHJT04} that was used to study selective CIs in \cite{LSST16} and again find favorable empirical performance for the HySI CIs.

The remainder of this paper is organized as follows.  Section \ref{sec:intuition} introduces the basic intuition for and construction of HySI CIs.  Section \ref{sec:framework} lays out the general model selection framework and assumptions under study involving an affine constraint on a vector of statistics.  Section \ref{sec:theory} details the general construction of HySI CIs and includes the main theoretical results of this paper on their correct uniform asymptotic coverage.  In Section \ref{sec:LASSO}, I show how the general framework of Sections \ref{sec:framework} and \ref{sec:theory} specialize to constructing a HySI CI for a regression coefficient after using the LASSO objective function to select the regression model of interest.  Sections \ref{sec:MC} and \ref{sec:empirical} present Monte Carlo simulation results and an empirical application for this post-LASSO model selection exercise.  Finally, Section \ref{sec:conclusion} concludes while proofs of the theoretical results in the paper are contained in a supplemental appendix.  All tables and figures are collected at the end of this document.

\section{Basic Ideas Behind the HySI Approach}\label{sec:intuition}

To impart intuition, I focus on a particular application for the construction of CIs after model selection, noting that a general approach that covers a wide range of applications is given in the following section.  In particular, consider the standard linear regression framework for which a response variable $y_i$ is modeled as a linear function of a predictor variable of interest $z_i$ and some subset of the control variables $X_{1i},\ldots,X_{pi}$ for $i=1,\ldots,n$.  Without imposing any assumptions about the true underlying relationship between the response, predictor of interest and controls, the researcher chooses a model $M\subset\{1,\ldots,p\}$ as the subset of indices corresponding to the controls of interest.  The researcher's target parameter of interest is equal to the population linear regression coefficient $\theta_M$ defined by
\[(\theta_M,\beta_{M})=\argmin_{\theta\in\mathbb{R},b_M\in\mathbb{R}^{|M|}}E\|y-z\theta-X_Mb_M\|^2,\]
where $y=(y_1,\ldots,y_n)^{\prime}$, $z=(z_1,\ldots,z_n)^{\prime}$ and $X_M$ is the submatrix of the design matrix $X=(x_1,\ldots,x_p)$ corresponding to model $M$ with $x_k=(X_{k1},\ldots,X_{kn})^{\prime}$.  

To establish the basic arguments, let us temporarily assume we have CIs with correct finite-sample coverage, regression coefficient estimators that are normally distributed in finite samples and that variance parameters are known.\footnote{These assumptions will all be relaxed to more realistic asymptotic assumptions in the following sections.} There is now a large literature enabling the construction of selective CIs with (asymptotically) correct coverage for a population regression coefficient conditional on the model $\widehat{M}\subset \{1,\ldots,p\}$ selected by the user for a variety of model selection criteria (e.g.,~\citealp{LSST16}; \citealp{TTLT16}; \citealp{TRTW18}).    That is, we have at our disposal a level $1-\alpha$ selective CI $CI_{\widehat M}^{S,\alpha}$ such that
\begin{equation}
\mathbb{P}\left(\theta_{\widehat M}\in CI_{\widehat M}^{S,\alpha}|\widehat M=M\right)\geq 1-\alpha \label{cond coverage}
\end{equation}
for all $M\subset\{1,\ldots,p\}$ (or some other relevant subset of the universe of models).  On the other hand, there is a growing literature enabling the construction of PoSI CIs with correct unconditional coverage for a regression coefficient chosen by \emph{any} model selection technique (e.g.,~\citealp{BBBZZ}; \citealp{KBBCGZ19}; \citealp{BPS20}).  That is, for any potentially data-dependent $\widehat{M}\subset \{1,\ldots,p\}$, we have at our disposal a level $1-\alpha$ PoSI CI $CI_{\widehat M}^{P,\alpha}$ such that
\begin{equation}
\mathbb{P}\left(\theta_{\widehat M}\in CI_{\widehat M}^{P,\alpha}\right)\geq 1-\alpha \label{POSI coverage}
\end{equation}
for all $\widehat M\subset\{1,\ldots,p\}$ (or some other relevant subset of the universe of models).

Selective CIs are typically constructed by expressing the model selection event $\{\widehat M=M\}$ in terms of a data-dependent truncation interval for the OLS estimator $\widehat\theta_{\widehat M}$ of $\theta_{\widehat M}$, where
\[(\widehat\theta_M,\widehat\beta_{M})=\argmin_{\theta\in\mathbb{R},b_M\in\mathbb{R}^{|M|}}\|y-z\theta-X_Mb_M\|^2.\]
This truncation interval depends upon a sufficient statistic $Z_{\widehat M}$ for the unknown nuisance parameter $\beta_{\widehat M}$ that is independent of $\widehat\theta_{\widehat M}$ after conditioning on the realization of $\widehat{M}$, i.e., $\{\widehat M=M\}=\{\widehat\theta_M\in [\mathcal{V}_M^-(Z_M),\mathcal{V}_M^+(Z_M)]\}$.\footnote{There is an additional element of the conditioning set $\mathcal{V}_M^0(Z_M)\geq 0$ that is suppressed in this section for simplicity of exposition.}  The typical construction of a selective CI then proceeds by invoking the fact that $\widehat\theta_{\widehat M}|\widehat M=M$ is distributed according to a normal distribution with mean $\theta_{M}$ truncated to the interval $[\mathcal{V}_M^-(Z_M),\mathcal{V}_M^+(Z_M)]$, and collecting all null hypothesized values of $\theta_{M}$ for which a test based upon this distribution evaluated at the realized value of $Z_M$ would fail to reject at level $\alpha$.  On the other hand, PoSI CIs typically take the form
\begin{equation*}
CI_{\widehat M}^{P,\alpha}=\widehat\theta_{\widehat M}\pm \sigma_{\widehat M}K_{\alpha}, 
\end{equation*}
where $\sigma_{M}$ is the standard deviation of $\widehat\theta_{M}$ and $K_{\alpha}$ is a constant that guarantees \eqref{POSI coverage} holds.  

My proposal in this context is to form a level $1-\alpha$ HySI CI $CI_{\widehat M}^{H,\alpha}$ that is constructed in analogy with the selective CI after modifying the conditioning event and appropriately adjusting the corresponding coverage level.  More specifically, this modified conditioning event is equal to the intersection of the model selection event expressed in terms of the sufficient statistic $Z_M$ and the (potentially false) event that $\theta_{\widehat M}$ lies inside of a level $1-\gamma>1-\alpha$ PoSI CI:
\begin{align*}
\left\{\widehat M=M\right\}\cap \left\{\theta_{\widehat M}\in CI_{\widehat M}^{P,\gamma}\right\}&=\left\{\widehat\theta_M\in [\mathcal{V}_M^-(Z_M),\mathcal{V}_M^+(Z_M)]\right\}\cap \left\{\widehat\theta_{\widehat M}\in [\theta_{\widehat M}-\sigma_{\widehat M}K_{\gamma},\theta_{\widehat M}+\sigma_{\widehat M}K_{\gamma}]\right\} \notag \\
&=\left\{\widehat\theta_{M}\in [\mathcal{V}_M^{-,H}(Z_M,\theta_M),\mathcal{V}_M^{+,H}(Z_M,\theta_M)]\right\}, 
\end{align*}
where 
\[\mathcal{V}_M^{-,H}(Z_M,\theta_M)=\max\left\{\mathcal{V}_M^-(Z_M),\text{ }\theta_M-\sigma_MK_{\gamma}\right\},\] 
\[\mathcal{V}_M^{+,H}(Z_M,\theta_M)=\min\left\{\mathcal{V}_M^+(Z_M),\text{ }\theta_M+\sigma_MK_{\gamma}\right\}\] 
(using the convention that $[a,b]=\emptyset$ if $b<a$).  A HySI CI is then constructed by invoking the fact that $\widehat\theta_{\widehat M}|\{\widehat M=M\}\cap \{\theta_{\widehat M}\in CI_{\widehat M}^{P,\gamma}\}$ is distributed according to a normal distribution with mean $\theta_{M}$ truncated to the interval $[\mathcal{V}_M^{-,H}(Z_M,\theta_M),\mathcal{V}_M^{+,H}(Z_M,\theta_M)]$.  It is defined as all null hypothesized values of $\theta_{M}$ for which a test based upon this distribution evaluated at the realized value of $Z_M$ would fail to reject at the adjusted level of $(\alpha-\gamma)/(1-\gamma)$.  

When $\mathbb{P}(\widehat M=M)$ is small, one (or both) of the bounds of the truncation interval used to form a selective CI $[\mathcal{V}_M^-(Z_M),\mathcal{V}_M^+(Z_M)]$ tends to be very close to $\widehat\theta_M$ so that a test based upon the distribution of $\widehat\theta_M|\{\widehat M=M\}$ fails to reject for many hypothesized values of the mean $\theta_M$ of $\widehat\theta_M$, including very large ones.  Since it is based upon inverting such a test, this induces the selective CI to become very long.  The additional condition $\theta_{\widehat M}\in CI_{\widehat M}^{P,\gamma}$ used to form the HySI CI bounds the value $\theta_M$ can take from above and below so that by conditioning on $\{\widehat\theta_{M}\in [\mathcal{V}_M^{-,H}(Z_M,\theta_M),\mathcal{V}_M^{+,H}(Z_M,\theta_M)]\}$ rather than $\{\widehat\theta_{M}\in [\mathcal{V}_M^{-}(Z_M,\theta_M),\mathcal{V}_M^{+}(Z_M,\theta_M)]\}$, the values of $\theta_M$ under consideration in the formation of the HySI CI are bounded above and below by $\widehat\theta_{M}\pm\sigma_MK_\gamma$.  In contrast to the selective CI, this fact implies that the level $1-\alpha$ HySI CI lies inside of the level $1-\gamma$ PoSI CI $CI_{\widehat M}^{P,\gamma}$ so that the former is never longer than the latter.  In addition, the latter CI is not much wider than the level $1-\alpha$ PoSI CI $CI_{\widehat M}^{P,\alpha}$ by virtue of the fact that the length of $CI_{\widehat M}^{P,\alpha}$ depends upon $\alpha$ as a constant times $\sqrt{\log (\alpha^{-1})}$.\footnote{This is shown by \cite{BBN18} under Gaussian errors with a fixed design matrix but the argument can be extended via the CLT (see \citealp{BPS20} and \citealp{KMB20}).}  On the other hand, when $\mathbb{P}(\widehat M=M)$ is large, the truncation interval used to form the selective CI tends to be very wide so that the distribution of $\widehat\theta_M|\{\widehat M=M\}$ is close to the unconditional distribution of $\widehat\theta_M$ and the selective CI is very close to its short naive counterpart.  Since $\mathcal{V}_M^-(Z_M)$ is large and negative and $\mathcal{V}_M^+(Z_M)$ is large and positive in this case, the HySI truncation interval $[\mathcal{V}_M^{-,H}(Z_M,\theta_M),\mathcal{V}_M^{+,H}(Z_M,\theta_M)]$ becomes very close to the selective truncation interval yielding a HySI CI that is very close to the $(1-\alpha)/(1-\gamma)$ selective and naive CIs.

The reason behind inverting tests at the adjusted level $(\alpha-\gamma)/(1-\gamma)$ (rather than $\alpha$) is to account for the fact that the modified conditioning event is not necessarily satisfied by a given realization of the data since $\mathbb{P}(\theta_{\widehat M}\in CI_{\widehat M}^{P,\gamma})<1$.  To see why this adjusted level yields correct unconditional coverage of the HySI CI, note that analogous arguments to those used to guarantee correct conditional coverage \eqref{cond coverage} can be used to guarantee
\begin{equation}
\mathbb{P}\left(\theta_{\widehat M}\in CI_{\widehat M}^{H,\alpha}\left|\widehat M=M,\theta_{\widehat M}\in CI_{\widehat M}^{P,\gamma}\right.\right)\geq \frac{1-\alpha}{1-\gamma} \label{hybrid cond coverage}
\end{equation}
for all $ M\subset\{1,\ldots,p\}$ so that
\begin{align*}
\mathbb{P}\left(\theta_{\widehat M}\in CI_{\widehat M}^{H,\alpha}\right)
&\geq \mathbb{P}\left(\theta_{\widehat M}\in CI_{\widehat M}^{H,\alpha}\left|\theta_{\widehat M}\in CI_{\widehat M}^{P,\gamma}\right.\right)\mathbb{P}\left(\theta_{\widehat M}\in CI_{\widehat M}^{P,\gamma}\right)\geq \frac{1-\alpha}{1-\gamma}(1-\gamma)=1-\alpha
\end{align*}
for all $\widehat M\subset\{1,\ldots,p\}$, where the final inequality follows from \eqref{POSI coverage}, \eqref{hybrid cond coverage} and the law of iterated expectations.

\section{General Asymptotic Model Selection Framework}\label{sec:framework}

In this section I introduce a set of very general assumptions to incorporate several forms of model selection and post-selection targets of inferential interest.  I discuss how these assumptions apply in many model selection settings.  In Section \ref{sec:LASSO}, I provide more detail about how these assumptions hold in the context of inference on a regression coefficient of interest after using LASSO to select controls variables.

Suppose we use a data set of $n$ observations that is realized from an unknown probability measure $\mathbb{P}\in\mathcal{P}_n$ to select a model $M$ from a finite set of models $\mathcal{M}=\{1,\ldots,|\mathcal{M}|\}$.  I require the set of probability measures $\mathcal{P}_n$ to satisfy a uniform version of the model selection condition of Markovic et al (2018). Letting $\widehat M_n$ denote the (random) model selected by the data, suppose that the event that a given model $M\in\mathcal{M}$ is selected is equivalent to a random vector $D_n(M)$ satisfying an affine constraint according to the following assumption.

\begin{assumption} \label{ass: selection contraint}
For all $M\in \mathcal{M}$, $\widehat{M}_n=M$ if and only if $A_M{D}_n(M)\leq \widehat a_{M,n}$, where $A_M$ is a fixed matrix and $\widehat a_{M,n}$ is a random vector such that for all $\varepsilon>0$, 
\[\lim_{n\rightarrow\infty}\sup_{\mathbb{P}\in \mathcal{P}_n}\mathbb{P}\left(\|\widehat a_{M,n}-a_{M,n}(\mathbb{P})\|>\varepsilon\right)=0\]
for some vector-valued sequence of functions $a_{M,n}(\mathbb{P})$ such that for some finite $\bar\lambda$, $\|a_{M,n}(\mathbb{P})\|\leq\bar\lambda$ for all $\mathbb{P}\in\mathcal{P}_n$ and $n\geq 1$.
\end{assumption}

Several papers in the selective inference literature have shown that the model selection event $\{\widehat{M}_n=M\}$ is equivalent to the event that an affine constraint holds on a lower-dimensional statistic $D_n(M)$ that is a function of the underlying data in accordance with Assumption \ref{ass: selection contraint}.  These affine constraints typically arise from Karush-Khun-Tucker (KKT) necessary and sufficient conditions from optimizing the objective function determining which model is selected, by comparing the values of different statistics across the steps of an iterative selection procedure and/or by comparing the values of statistics evaluated at different tuning parameters when model selection tuning parameters are data-dependent.  For example, \cite{LSST16} show that for a fixed LASSO penalty parameter $\lambda$, the LASSO model selection event characterizing the set of non-zero regression coefficients and their signs is equivalent to $\{A_M{D}_n(M)\leq \widehat a_{M,n}\}$ for which $A_M$ is a matrix of zeros, ones and negative ones, $D_n(M)$ is a vector of scaled least squares regression estimates and inner products of regressors and residuals and $\widehat a_{M,n}$ is a vector that is a function of $\lambda$ and the design matrix.  This vector $\widehat a_{M,n}$ naturally satisfies the uniform convergence in probability in the assumption by a uniform law of large numbers and standard moments conditions on the underlying design matrix.  

As another example, \cite{TTLT16} show that for a fixed number of steps along the FS and LAR algorithms, the selection event characterizing the active regressors and their signs is equivalent to $\{A_M{D}_n(M)\leq \widehat a_{M,n}\}$ for which $A_M$ is a matrix of ones and negative ones, $D_n(M)$ is a vector with elements that are functions of inner products of orthognally projected regressors and dependent variables and $\widehat a_{M,n}=0$.  Using the equivalence between the solution path of LASSO and a modified version of LAR (\citealp{EHJT04}), \cite{TTLT16} provide a similar affine characterization for models chosen along the solution path of LASSO (rather than for a fixed LASSO penalty parameter).  Finally, \cite{MXT18} and \cite{TT18} show the affine characterization of the selection event holds for various selection procedures when random noise is added to the selection criterion and/or the selection criterion uses a data-dependent tuning parameter (such as LASSO with $\lambda$ chosen via cross-validation).  Describing the various quantities for these latter cases is rather involved and I refer the interested reader to \cite{MXT18} and \cite{TT18}.

I am interested in constructing a CI for a scalar parameter that is chosen based upon the selected model $\widehat{M}_n$.  I denote this target parameter as $\mu_{T,n}(\widehat{M}_n)$.\footnote{To streamline notation I slightly abuse it by using $\mu_{T,n}(M)$ to denote $\mu_{T,n}(M;\mathbb{P})$, the $M^{th}$ element of $\mu_{T,n}(\mathbb{P})$ defined in Assumption \ref{ass: joint normality}.}  For any $M$, assume that in the absence of data-dependent selection, there is an asymptotically Gaussian statistic $T_n(M)$ centered around $\mu_{T,n}({M})$.  Further assume that the full vectors of statistics $T_n(M)$ and the statistics determining selection $D_n(M)$ are uniformly jointly asymptotically normal under $\mathbb{P}\in \mathcal{P}_n$ with centering vectors $\mu_{T,n}$ and $\mu_{D,n}$ and limiting covariance matrix $\Sigma$ that may depend upon $\mathbb{P}$.  I use the following strengthening of \cite{MXT18} to full joint convergence of all statistics because I focus on unconditional inferential statements that do not condition on the selected model.  For any matrix $A$, let $\lambda_{\min}(A)$ and $\lambda_{\max}(A)$ denote its minimum and maximum eigenvalues.

\begin{assumption} \label{ass: joint normality}
For $T_n=(T_n(1),\ldots,T_n(|\mathcal{M}|))^{\prime}\in\mathbb{R}^{|\mathcal{M}|}$ and $D_n=(D_n(1)^{\prime},\ldots,D_n(|\mathcal{M}|)^{\prime})^{\prime}$ and the class of Lipschitz functions that are bounded in absolute value by one and have Lipschitz constant bounded by one, $BL_1$, there exist sequences of functions $\mu_{T,n}(\mathbb{P})$ and $\mu_{D,n}(\mathbb{P})$ and a function $\Sigma(\mathbb{P})$ such that for $(T_{\mathbb{P}}^{*\prime},D_{\mathbb{P}}^{*\prime})^{\prime}\sim\mathcal{N}(0,\Sigma(\mathbb{P}))$ with
\[\Sigma=\left(\begin{array}{cc}
\Sigma_T & \Sigma_{TD} \\
\Sigma_{DT} & \Sigma_D
\end{array}\right),\]
\[\lim_{n\rightarrow\infty}\sup_{\mathbb{P}\in \mathcal{P}_n}\sup_{f\in BL_1}\left|\mathbb{E}_{\mathbb{P}}\left[f\left(\begin{array}{c}
T_n-\mu_{T,n}(\mathbb{P}) \\
D_n-\mu_{D,n}(\mathbb{P})
\end{array}\right)\right]-\mathbb{E}_{\mathbb{P}}\left[f\left(\begin{array}{c}
T_{\mathbb{P}}^{*} \\
D_{\mathbb{P}}^{*}
\end{array}\right)\right]\right|=0.\]
Furthermore, for some finite $\bar\lambda>0$, $1/\bar \lambda\leq \Sigma_T(M,M;\mathbb{P})\leq \bar\lambda$ and $1/\bar \lambda\leq\lambda_{\min}(\Sigma_D^{(M)}(\mathbb{P}))\leq\lambda_{\max}(\Sigma_D^{(M)}(\mathbb{P}))\leq \bar\lambda$ for all $M\in\mathcal{M}$ and $\mathbb{P}\in \mathcal{P}_n$, where $\Sigma_D^{(M)}$ is the covariance matrix of $D^*(M)$. 
\end{assumption} 

The statistics $T_n(M)$ typically take the form of scaled (linear functionals of) sample regression coefficient estimates under the imposed model $M$ with $\mu_{T,n}(M)$ being equal to the scaled (linear functionals of) population regression counterparts.  As discussed following Assumption \ref{ass: selection contraint}, the statistics $D_n(M)$ typically take the form of scaled sample regression parameter estimates and functions of inner products of dependent variables, regressors and residuals (with the additon of noise terms for applications involving randomization).  Thus, $T_n$ and $D_n$ naturally satisfy joint uniform CLTs under standard assumptions on the data.  Indeed, \cite{MXT18} verify marginal CLTs on $(T_n(M),D_N(M)^{\prime})^\prime$ for a given $M$ for inference on regression coefficients after LASSO model selection with a fixed penalty parameter and a penalty parameter chosen via randomized cross-validation and after randomized information criteria-based selection.  It is straightforward to extend these results to show that the joint uniform CLT in Assumption \ref{ass: joint normality} holds for these problems.  Similarly, \cite{TRTW18} show that both linear functionals of sample regression coefficient estimates and the selection events involved in models selected for a fixed number of steps along the FS and LAR algorithms and along the solution path of LASSO are functions of a ``master statistic'', enabling a joint uniform CLT of the form given in Assumption \ref{ass: joint normality}.\footnote{Although these latter results impose a fixed design matrix, they easily extend to random designs under suitable assumptions.}  It is important to note that Assumption \ref{ass: joint normality} does not impose eigenvalue bounds on the full covariance matrix $\Sigma$, allowing this matrix to be singular.

In order to form asymptotically valid HySI CIs, I require the use of uniformly consistent estimators $\widehat\Sigma_{T,n}$ and $\widehat\Sigma_{DT,n}$ for the covariance matrices in Assumption \ref{ass: joint normality}.  Let $\|\cdot\|$ denote the Frobenius norm.

\begin{assumption} \label{ass: covariance estimator}
There exist estimators $\widehat\Sigma_{T,n}$ and $\widehat\Sigma_{DT,n}$ such that for all $\varepsilon>0$,
\[\lim_{n\rightarrow\infty}\sup_{\mathbb{P}\in \mathcal{P}_n}\mathbb{P}\left(\|\widehat\Sigma_{T,n}-\Sigma_T(\mathbb{P})\|>\varepsilon\right)=0 \text{ and } \lim_{n\rightarrow\infty}\sup_{\mathbb{P}\in \mathcal{P}_n}\mathbb{P}\left(\|\widehat\Sigma_{DT,n}-\Sigma_{DT}(\mathbb{P})\|>\varepsilon\right)=0.\]
\end{assumption}

Unlike the PoSI CIs of \cite{BPS20}, using an inconsistent ``conservative'' estimator of $\Sigma_T$ that consistently overestimates its diagonal values will not lead to HySI CIs with correct asymptotic coverage.  As will become more apparent in the following section detailing the HySI CI construction, consistent estimation of $\Sigma_T$ and $\Sigma_{DT}$ is crucial to forming a random vector $Z_{M,n}$ that is asymptotically independent of the statistic $T_n(M)$.  In the context of selection in the linear regression model, the (co)variances $\Sigma_T$ and $\Sigma_{DT}$ are possible to consistently estimate when (i) the regressors are either random or constant (corresponding to an intercept term) or (ii) the ``true'' model is known a priori.  The former has been shown by, e.g.,~\cite{KBB18} and references therein and the latter is well known.  Although much of the selective inference and PoSI literature has focused on the case of a fixed design matrix, case (i) is arguably more relevant for most practical applications.  For typical applications, \cite{KBB18} show how the elements of $\Sigma_T$ and $\Sigma_{DT}$ can be estimated consistently using standard heteroskedasticity-robust methods in linear regression contexts with iid data under standard moment conditions, applying to both homoskedastic and heteroskedastic data.  \cite{Fre81} and \cite{BBBGPTZ16} show how the elements of $\Sigma_T$ and $\Sigma_{DT}$ can be estimated via pairs bootstrap (see also \citealp{MXT18}).  The consistency arguments can be strengthened to the uniform consistency requirement of Assumption \ref{ass: covariance estimator} straightforwardly.


I make one final high-level assumption on the existence of a PoSI CI with correct unconditional uniform asymptotic coverage of the parameter of interest $\mu_{T,n}(\widehat{M}_n)$. 

\begin{assumption} \label{ass: PoSI CS}
For any $\alpha\in (0,1)$, we have a CI of the form 
\[CI_{n,\widehat{M}_n}^{P,\alpha}=T_n(\widehat{M}_n)\pm \sqrt{\widehat \Sigma_{T,n}(\widehat{M}_n,\widehat{M}_n)}K_{n,\alpha}\]
that satisfies $\liminf_{n\rightarrow \infty}\inf_{\mathbb{P}\in\mathcal{P}_n}\mathbb{P}\left(\mu_{T,n}(\widehat{M}_n;\mathbb{P})\in CI_{n,\widehat{M}_n}^{P,\alpha}\right)\geq 1-\alpha$ and for any $\varepsilon>0$
\[\lim_{n\rightarrow\infty}\sup_{\mathbb{P}\in \mathcal{P}_n}\mathbb{P}\left(|K_{n,\alpha}-K_{\alpha}(\mathbb{P})|>\varepsilon\right)=0\]
for some function $K_{\alpha}(\mathbb{P})$ such that for some finite $\bar\lambda$, $0\leq K_{\alpha}(\mathbb{P})\leq\bar\lambda$ for all $\mathbb{P}\in\mathcal{P}_n$.
\end{assumption}

For inference on (linear functionals of) population regression coefficients after model selection in the linear regression framework, $K_{n,\alpha}$ typically takes the form of (an upper bound on) the $(1-\alpha)$-quantile of the maximum of a sequence of correlated standard normal random variables, for which the correlation matrix is derived from $\widehat\Sigma_{T,n}$, or an asymptotically equivalent bootstrap version.  The uniformly consistent estimation of $\widehat\Sigma_{T,n}$ implied by Assumption \ref{ass: covariance estimator} then immediately implies the uniform consistency of $K_{n,\alpha}$ for $K_\alpha$ required by Assumption \ref{ass: PoSI CS} while the results of \cite{BPS20} imply the uniform coverage requirement of the assumption for several examples of PoSI CIs.

\section{HySI CIs and Uniform Asymptotic Validity}\label{sec:theory}

We are now equipped with the ingredients needed to define the $(1-\alpha)$-level HySI CI, $CI_{n,\widehat{M}_n}^{H,\alpha}$, for $\mu_{T,n}(\widehat{M}_n)$.  To begin describing the HySI CI construction, it is useful to express the conditioning event in Assumption \ref{ass: selection contraint} in terms of a data-dependent interval for the target statistic $T_n(\widehat M_n)$.  The bounds of this interval are expressed in terms of a directly-computable random vector $Z_{M,n}$ that is asymptotically independent of $T_n(M)$:
\[Z_{M,n}=D_n(M)-\left(\widehat \Sigma_{DT,n}^{(M)}/\widehat\Sigma_{T,n}(M,M)\right)T_n(M),\]
where $\widehat \Sigma_{DT,n}^{(M)}$ is the estimated covariance vector between $T_n(M)$ and $D_n(M)$.  The following lemma follows from a slight extension of the arguments used to prove Lemma 5.1 in \cite{LSST16}.

\begin{lemma} \label{lem: truncation interval}
Under Assumption \ref{ass: selection contraint}, the conditioning set for any model $M\in \mathcal{M}$ being selected can be expressed as follows:
\[\left\{\widehat M_n=M\right\}=\left\{\mathcal{V}_{M,n}^{-}(Z_{ M,n})\leq T_n(M)\leq \mathcal{V}_{M,n}^{+}(Z_{ M,n}),\mathcal{V}_{M,n}^{0}(Z_{ M,n})\geq 0\right\},\]
where 
\begin{align*}
\mathcal{V}_{M,n}^{-}(z)&=\max_{j:(A_M\widehat \Sigma_{DT,n}^{(M)}/\widehat\Sigma_{T,n}(M,M))_j<0}\frac{\widehat a_{M,n,j}-(A_Mz)_j}{(A_M\widehat \Sigma_{DT,n}^{(M)}/\widehat\Sigma_{T,n}(M,M))_j} \\
\mathcal{V}_{M,n}^{+}(z)&=\min_{j:(A_M\widehat \Sigma_{DT,n}^{(M)}/\widehat\Sigma_{T,n}(M,M))_j>0}\frac{\widehat a_{M,n,j}-(A_Mz)_j}{(A_M\widehat \Sigma_{DT,n}^{(M)}/\widehat\Sigma_{T,n}(M,M))_j} \\
\mathcal{V}_{M,n}^{0}(z)&=\min_{j:(A_M\widehat \Sigma_{DT,n}^{(M)}/\widehat\Sigma_{T,n}(M,M))_j=0}\widehat a_{M,n,j}-(A_Mz)_j.
\end{align*}
\end{lemma} 
The HySI CI is constructed from the distribution function of $T_n(\widehat{M}_n)$ after conditioning on the events $\{\widehat{M}_n=M\}$ and 
\[\left\{\mu_{T,n}(\widehat{M}_n)\in CI_{n,\widehat{M}_n}^{P,\gamma}\right\}\]
\[=\left\{\mu_{T,n}(\widehat{M}_n)- \sqrt{\widehat \Sigma_{T,n}(\widehat{M}_n,\widehat{M}_n)}K_{n,\gamma}\leq T_n(\widehat{M}_n)\leq \mu_{T,n}(\widehat{M}_n)+ \sqrt{\widehat \Sigma_{T,n}(\widehat{M}_n,\widehat{M}_n)}K_{n,\gamma}\right\}\] 
(by Assumption \ref{ass: PoSI CS}) for some $\gamma\in(0,\alpha)$.  More specifically, let $F_{TN}(\cdot;\mu,\sigma^2,\mathcal{L},\mathcal{U})$ denote the truncated normal distribution function of $\xi|\{\mathcal{L}\leq \xi\leq \mathcal{U}\}$ for $\xi\sim\mathcal{N}(\mu,\sigma^2)$.  For $\alpha\in(0,1)$, define $\widehat\mu_{T,n}^{H,\alpha}(\widehat{M}_n)$ to solve 
\[F_{TN}\left(T_n(\widehat{M}_n);\mu,\widehat\Sigma_{T,n}(\widehat{M}_n,\widehat{M}_n),\mathcal{V}_{\widehat M_n,n}^{-,H}(Z_{\widehat M_n,n},\mu),\mathcal{V}_{\widehat M_n,n}^{+,H}(Z_{\widehat M_n,n},\mu)\right)=1-\alpha\] 
in $\mu$, where 
\[\mathcal{V}_{M,n}^{-,H}(z,\mu)=\max\left\{\mathcal{V}_{M,n}^{-}(z),\text{ }\mu-\sqrt{\widehat\Sigma_{T,n}(M,M)}K_{n,\gamma}\right\},\] 
\[\mathcal{V}_{M,n}^{+,H}(z,\mu)=\min\left\{\mathcal{V}_{M,n}^{+}(z),\text{ }\mu+\sqrt{\widehat\Sigma_{T,n}(M,M)}K_{n,\gamma}\right\}.\]
In turn, $CI_{n,\widehat{M}_n}^{H,\alpha}$ is defined as 
\begin{equation}
CI_{n,\widehat{M}_n}^{H,\alpha}=\left[\widehat\mu_{T,n}^{H,\frac{\alpha-\gamma}{2(1-\gamma)}}(\widehat{M}_n),\widehat\mu_{T,n}^{H,1-\frac{\alpha-\gamma}{2(1-\gamma)}}(\widehat{M}_n)\right], \label{eq: hyb CI}
\end{equation}
where $\widehat\mu_{T,n}^{H,\frac{\alpha-\gamma}{2(1-\gamma)}}(\widehat{M}_n)$ and $\widehat\mu_{T,n}^{H,1-\frac{\alpha-\gamma}{2(1-\gamma)}}(\widehat{M}_n)$ are used instead of $\widehat\mu_{T,n}^{H,\alpha/2}(\widehat{M}_n)$ and $\widehat\mu_{T,n}^{H,1-\alpha/2}(\widehat{M}_n)$ to account for the fact that the probability of the conditioning event $\{\mu_{T,n}(\widehat{M}_n)\in CI_{n,\widehat{M}_n}^{P,\gamma}\}$ is only bounded below by $1-\gamma$ under all sequences of probability measures $\{\mathbb{P}_n\}$ (by Assumption \ref{ass: PoSI CS}).  For simplicity, I focus on the two-sided equal-tailed version of the HySI CI as defined in \eqref{eq: hyb CI} but note that the uniform asymptotic validity results presented here also apply to one-sided and non-equal-tailed versions for which $\widehat\mu_{T,n}^{H,\frac{\alpha-\gamma}{2(1-\gamma)}}(\widehat{M}_n)$ and $\widehat\mu_{T,n}^{H,1-\frac{\alpha-\gamma}{2(1-\gamma)}}(\widehat{M}_n)$ are replaced by any $\widehat\mu_{T,n}^{H,q_1}(\widehat{M}_n)$ and $\widehat\mu_{T,n}^{H,1-q_2}(\widehat{M}_n)$ such that $q_1+q_2=(\alpha-\gamma)/(1-\gamma)$.  

Once the PoSI constant $K_{n,\gamma}$ is found, construction of the HySI CI is computationally straightforward since it just involves finding the zeros of two continuous functions.  Since $K_{n,\gamma}$ must be computed to form the HySI CI, this implies that the HySI and PoSI CIs share approximately the same degree of computational complexity.  \cite{BBBZZ} provide code for efficiently computing non-conservative $K_{n,\gamma}$ values (in the sense that the coverage requirement in Assumption \ref{ass: PoSI CS} holds with equality) in linear regression contexts when 20 or less covariates are subjected to model selection.  \cite{BBBZZ} and \cite{BPS20} also discuss computationally straightforward methods for computing conservative values of $K_{n,\gamma}$ that satisfy Assumption \ref{ass: PoSI CS} even when the number of models under consideration is very large by appealing to bounds on the quantiles of the maximum of correlated Gaussian random variables.

I now state a result establishing the asymptotic coverage of $CI_{n,\widehat{M}_n}^{H,\alpha}$ conditional on the realization of the selected model $\widehat{M}_n$ and the possibly false event $\{\mu_{T,n}(\widehat{M}_n)\in CI_{n,\widehat{M}_n}^{P,\gamma}\}$.

\begin{proposition}\label{prop: cond coverage}
Under Assumptions \ref{ass: selection contraint}--\ref{ass: PoSI CS},
\begin{align*}
&\lim_{n\rightarrow\infty}\sup_{\mathbb{P}\in\mathcal{P}_n}\left|\mathbb{P}\left(\mu_{T,n}(\widehat{M}_n)\in CI_{n,\widehat{M}_n}^{H,\alpha}\left|\widehat{M}_n=M,\mu_{T,n}(\widehat{M}_n)\in CI_{n,\widehat{M}_n}^{P,\gamma}\right.\right)-\frac{1-\alpha}{1-\gamma}\right| \\
&\qquad \qquad \times\mathbb{P}\left(\widehat{M}_n=M,\mu_{T,n}(\widehat{M}_n)\in CI_{n,\widehat{M}_n}^{P,\gamma}\right)=0
\end{align*}
for all $M\in \mathcal{M}$.
\end{proposition}

Using the results from Proposition \ref{prop: cond coverage}, we can show that $CI_{n,\widehat{M}_n}^{H,\alpha}$ has correct unconditional coverage at level $1-\alpha$ and a controlled degree of nonsimilarity.  This is the main theoretical result of the paper.

\begin{proposition}\label{prop: uncond coverage}
Under Assumptions \ref{ass: selection contraint}--\ref{ass: PoSI CS},
\[\liminf_{n\rightarrow\infty}\inf_{\mathbb{P}\in\mathcal{P}_n}\mathbb{P}\left(\mu_{T,n}(\widehat{M}_n;\mathbb{P})\in CI_{n,\widehat{M}_n}^{H,\alpha}\right)\geq 1-\alpha\]
and
\[\limsup_{n\rightarrow\infty}\sup_{\mathbb{P}\in\mathcal{P}_n}\mathbb{P}\left(\mu_{T,n}(\widehat{M}_n;\mathbb{P})\in CI_{n,\widehat{M}_n}^{H,\alpha}\right)\leq \frac{1-\alpha}{1-\gamma}.\]
\end{proposition}

 The user is free to choose any value of $\gamma$ with $0\leq\gamma\leq\alpha$ in the construction of the HySI CI.  There is no value of $\gamma$ that is optimal uniformly across the parameter space in terms of CI length measures.  Rather, as we can see from the lower and upper bounds given in the above proposition, $\gamma$ controls the degree of (asymptotic) non-similarity of the HySI CI with the CI being closer to (asymptotically) similar when $\gamma$ is small.\footnote{A similar CI is defined as having identical coverage probability uniformly across the parameter space.}  Similar CIs are not necessarily desirable in this context.  In fact when $\gamma=0$, the HySI CI is identical to the (similar) level $1-\alpha$ selective CI.  On the other hand, for $\gamma=\alpha$, the HySI CI is equal to the (non-similar) level $1-\gamma$ PoSI CI.  

 The choice of $\gamma$ trades off the length properties of the HySI CI over different realizations of the data.  I recommend a small but non-negligible value of $\gamma$ such as $\gamma=\alpha/10$ to attain a CI that is not much longer than the selective CI when the model selected by the data is selected with high probability (so that the selective CI is short) without compromising a lot of length when this does not occur.  Proposition 3 of \cite{AKM20} implies that when a given model is selected with probability approaching one, the $(1-\alpha)$--level HySI CI converges to a $(1-\alpha)/(1-\gamma)$--level selective CI which in turn converges to a $(1-\alpha)/(1-\gamma)$--level naive CI, where $\gamma\in(0,\alpha)$ is chosen by the user.  When using the recommended value of $\gamma=\alpha/10$ to construct the HySI CI, this means that $99\%$, $95\%$ and $90\%$ HySI CIs converge to $99.1\%$, $95.5\%$ and $90.9\%$ naive CIs. See Section \ref{sec:MC} and \cite{AKM20b} and \cite{AKM20} for further evidence that this choice works well in practice in related contexts.

\section{Application to Inference After LASSO Model Selection}\label{sec:LASSO}

I now specialize the general framework to the problem of constructing a HySI CI for a regression coefficient of interest after using LASSO to determine which covariates enter the regression model, dropping the simplifying assumptions of Section \ref{sec:intuition}.  Formally, suppose we have data $(z,X,y)\in\mathbb{R}^n\times \mathbb{R}^{n\times p}\times \mathbb{R}^n$ for which the rows of $y$ and $z$ are identically distributed random variables and the rows of $X$ are either identically distributed random vectors or have entries equal to one (corresponding to an intercept term).\footnote{In this and the following section, the rows of an arbitrary matrix or vector $B$ are denoted as $B_i$.}  We are interested in the population regression coefficient corresponding to the predictor of interest $z$ after selecting which of the control variables in $X$ should enter the regression model according to the non-zero subset of the vector $\widehat\beta$, where
\[\widehat\beta=\argmin_{\beta\in\mathbb{R}^{p}}\frac{1}{2}\|y^*-X^*\beta\|_2^2+\lambda\|\beta\|_1\]
with $y^*=(I-P_z)y$ and $X^*=(I-P_z)X$ for $P_z=zz^\prime/z^\prime z$ and $\lambda$ being the LASSO penalty parameter.  Letting $\widehat E_n$ denote the set of non-zero coefficients of $\widehat\beta$, we can characterize a model $M$ as a set of LASSO-selected controls $E$ and the sign of the LASSO regression coefficients corresponding to the selected controls $s_E$.  In other words, a given model $M$ is defined as a tuple $(E,s_E)$.  

Using the Karush-Khun-Tucker conditions for optimizing the LASSO objective function, \cite{LSST16} show that $\widehat M_n=(\widehat E_n,\mathrm{sign}(\widehat\beta_E))=(E,s_E)=M$ if and only if $A_MD_n(M)\leq \widehat a_{M,n}$, where
\[A_M=\left(\begin{array}{cc}
-\mathrm{diag}(s_E) & 0 \\
0 & I_{p-|E|} \\
0 & -I_{p-|E|}
\end{array}\right),\]
\[D_n(M)=\left(\begin{array}{c}
\sqrt{n}(X_E^{*\prime}X_E^{*})^{-1}X_E^{*\prime}y^* \\
n^{-1/2}X_{-E}^{*\prime}(y^*-X_{E}^{*}(X_E^{*\prime}X_E^{*})^{-1}X_E^{*\prime}y^*)
\end{array}\right),\]
\[\widehat a_{M,n}=\left(\begin{array}{c}
-\lambda\sqrt{n}\mathrm{diag}(s_E)(X_E^{*\prime}X_E^{*})^{-1}s_E \\
\lambda n^{-1/2}1_{p-|E|}-\lambda n^{-1/2}X_{-E}^{*\prime}X_{E}^{*}(X_E^{*\prime}X_E^{*})^{-1}s_E \\
\lambda n^{-1/2}1_{p-|E|}+\lambda n^{-1/2}X_{-E}^{*\prime}X_{E}^{*}(X_E^{*\prime}X_E^{*})^{-1}s_E
\end{array}\right),\]
with $X_E^*$ equal to the submatrix of $X^*$ composed of the columns of $X^*$ corresponding to $E$ and $X_{-E}^*$ equal to the submatrix of $X^*$ composed of the remaining columns.  Let $\tilde X=X-z(\mathbb{E}_{\mathbb{P}}[z'z])^{-1}\mathbb{E}_{\mathbb{P}}[z'X]$ and let $\tilde X_{E}^{\prime}$ denote the submatrix of $\tilde X$ composed of the columns of $\tilde X$  corresponding to $E$ and $\tilde X_{-E}^{\prime}$ denote the submatrix of $\tilde X$ composed of the remaining columns.  Assumption \ref{ass: selection contraint} thus holds with
\[a_{M,n}(\mathbb{P})=\left(\begin{array}{c}
-\lambda n^{-1/2}\mathrm{diag}(s_E)(\mathbb{E}_{\mathbb{P}}[\tilde X_{E,i}\tilde X_{E,i}^{\prime}])^{-1}s_E \\
\lambda n^{-1/2}1_{p-|E|}-\lambda n^{-1/2} \mathbb{E}_{\mathbb{P}}[\tilde X_{-E,i}\tilde X_{E,i}^{\prime}](\mathbb{E}_{\mathbb{P}}[\tilde X_{E,i}\tilde X_{E,i}^{\prime}])^{-1}s_E \\
\lambda n^{-1/2}1_{p-|E|}+\lambda n^{-1/2} \mathbb{E}_{\mathbb{P}}[\tilde X_{-E,i}\tilde X_{E,i}^{\prime}](\mathbb{E}_{\mathbb{P}}[\tilde X_{E,i}\tilde X_{E,i}^{\prime}])^{-1}s_E
\end{array}\right)\]
under standard moment, stationarity and dependence conditions on $\mathcal{P}_n$ that imply a uniform law of large numbers for $\widehat a_{M,n}$ and uniform moment bounds on $\mathbb{E}_{\mathbb{P}}[\tilde X_{-E,i}\tilde X_{E,i}^{\prime}]$ and $\mathbb{E}_{\mathbb{P}}[\tilde X_{E,i}\tilde X_{E,i}^{\prime}]$.

Letting $W_E=(z,X_E)$ and $e_1$ denote the first standard basis vector, we are interested in forming a CI that covers the (scaled) population regression coefficient on $z$ in the selected model as the target parameter
\[\mu_{T,n}(\mathbb{P},M)=\sqrt{n}e_1^{\prime}(\mathbb{E}_{\mathbb{P}}[W_{E,i}W_{E,i}^{\prime}])^{-1}\mathbb{E}_{\mathbb{P}}[W_{E,i}y_i]\]
for $E={\widehat E_n}$ using the corresponding sample regression coefficient 
\[T_n(M)=\sqrt{n}e_1'(W_E'W_E)^{-1}W_E'y\]
as a statistic.\footnote{CIs for unscaled population regression coefficients are formed by simply dividing the CIs for $\mu_{T,n}(\mathbb{P},M)$ by $\sqrt{n}$.}  With these definitions in mind, as well as
\[\mu_{D,n}(\mathbb{P},M)=\left(\begin{array}{c}
\sqrt{n}(\mathbb{E}_{\mathbb{P}}[\tilde X_{E,i}\tilde X_{E,i}'])^{-1}\mathbb{E}_{\mathbb{P}}[\tilde X_{E,i}\tilde y_i] \\
\sqrt{n}(\mathbb{E}_{\mathbb{P}}[\tilde X_{-E,i}\tilde y_i]-\mathbb{E}_{\mathbb{P}}[\tilde X_{-E,i}\tilde X_{E,i}'](\mathbb{E}_{\mathbb{P}}[\tilde X_{E,i}\tilde X_{E,i}'])^{-1}\mathbb{E}_{\mathbb{P}}[\tilde X_{E,i}\tilde y_i])
\end{array}\right),\]
Assumption \ref{ass: joint normality} holds under standard moment, stationarity and dependence conditions on $\mathcal{P}_n$ that imply a multivariate uniform CLT for the vector $(T_n',D_n')'$.  For Assumption \ref{ass: covariance estimator}, consider the  heteroskedasticity-robust estimators $\widehat{\Sigma}_{T,n}$ and $\widehat{\Sigma}_{DT,n}$ for which
\[\widehat{\Sigma}_{T,n}(M,M^{\prime})=e_1^{\prime}\left(\frac{1}{n}\sum_{i=1}^nW_{E,i}W_{E,i}^{\prime}\right)^{-1}\left(\frac{1}{n}\sum_{i=1}^nW_{E,i}W_{E^{\prime},i}^{\prime}\hat u_{E,i}\hat u_{E^{\prime},i}\right)\left(\frac{1}{n}\sum_{i=1}^nW_{E^\prime,i}W_{E^\prime,i}^{\prime}\right)^{-1}e_1,\]
\begin{gather*}
\widehat{\Sigma}_{DT,n}((M-1)p+1:Mp,M^\prime) \\
=\left(
\begin{array}{cc}
\left(\frac{1}{n}\sum_{i=1}^nX_{E,i}^*X_{E,i}^{*\prime}\right)^{-1} & 0 \\
0 & I
\end{array}\right)\left(\frac{1}{n}\sum_{i=1}^n\left(
\begin{array}{c}
X_{E,i}^* \\
X_{-E,i}^*
\end{array}\right)W_{E^\prime,i}^\prime u_{E,i}^*\hat u_{E^\prime,i}\right)\left(\frac{1}{n}\sum_{i=1}^nW_{E^\prime,i}W_{E^\prime,i}^{\prime}\right)^{-1}e_1
\end{gather*}
where $\hat u_{E,i}=y_i-W_{E,i}^{\prime}\widehat\beta_{M,n}$ and $u_{E,i}^*=y_i^*-X_{E,i}^*\widehat\beta_{M,n}^*$ with $\widehat\beta_{M,n}=(W_E^\prime W_E)^{-1}W_E^\prime y$ and $\widehat\beta_{M,n}^*=(X_{E}^{*\prime}X_{E}^{*})^{-1}X_{E}^{*\prime} y^*$.  Slight extensions of the arguments in \cite{KBB18} from pointwise to uniform consistency provide that Assumption \ref{ass: covariance estimator} holds for $\widehat{\Sigma}_{T,n}$ and $\widehat{\Sigma}_{DT,n}$ when the data are independent under standard moment conditions on $\mathcal{P}_n$.

Finally, Assumption \ref{ass: PoSI CS} holds by the results of \cite{BPS20} when using one of the PoSI CIs discussed in that paper.  In particular, let $K_{n,\alpha}$ equal the $(1-\alpha)$-quantile of
\[\max_i |Z_i| \text{ for } Z\sim\mathcal{N}(0,\Omega)\]
with $\Omega=\mathrm{corr}(\widehat \Sigma_{T,n})\equiv \mathrm{diag}(\widehat \Sigma_{T,n})^{\dagger/2}\widehat \Sigma_{T,n}\mathrm{diag}(\widehat \Sigma_{T,n})^{\dagger/2}$, where $A^{\dagger}$ denotes the Moore-Penrose inverse of matrix $A$ and $A^{1/2}$ denotes the symmetric nonnegative definite square root of a symmetric nonnegative definite matrix $A$.  By Assumption \ref{ass: covariance estimator}, 
\[\lim_{n\rightarrow\infty}\sup_{\mathbb{P}\in \mathcal{P}_n}\mathbb{P}\left(|K_{n,\alpha}-K_{\alpha}(\mathbb{P})|>\varepsilon\right)=0\]
for any $\varepsilon>0$, where $K_{\alpha}(\mathbb{P})$ is equal to the $(1-\alpha)$-quantile of $\max_i |Z_i| \text{ for } Z\sim\mathcal{N}(0,\Omega)$ with $\Omega=\mathrm{corr}(\Sigma_T(\mathbb{P}))$.  For $\alpha\neq 1$, $0\leq K_{\alpha}(\mathbb{P})\leq\bar\lambda$ for some finite $\bar\lambda$ and any probability measure $\mathbb{P}$.  Theorem 2.3 of \cite{BPS20} provides sufficient conditions on $\mathcal{P}_n$ that imply 
$\liminf_{n\rightarrow \infty}\inf_{\mathbb{P}\in\mathcal{P}_n}\mathbb{P}\left(\mu_{T,n}(\widehat{M}_n;\mathbb{P})\in CI_{n,\widehat{M}_n}^{P,\alpha}\right)\geq 1-\alpha$ for $CI_{n,\widehat{M}_n}^{P,\alpha}$ formed according to Assumption \ref{ass: PoSI CS}.\footnote{Note that the form of PoSI CI introduced here is a less conservaitve version that incorporates the fact that the predictor of interest $z$ is protected from variable selection, referred to as ``PoSI1'' by \cite{BBBZZ}.}

\section{Finite-Sample Properties of Confidence Intervals}\label{sec:MC}

In order to investigate the finite-sample properties of HySI CIs and compare them to existing CIs in a variety of settings, I examine Monte Carlo experiments for the application described in the previous section.  The DGPs I study in these Monte Carlo experiments are designed to closely match those studied in the simulations of \cite{TRTW18} and \cite{BPS20} who focus on the different application of inference on the variable selected across the steps of the LAR algorithm.  More specifically, I consider data generated from the standard linear regression model 
\begin{equation}
y=\theta z+X\beta+u \label{reg eqn}
\end{equation}
where $y$ is an $n\times 1$ vector of observations of the outcome of interest, $\theta\in\mathbb{R}$, $z$ is an $n\times 1$ vector of observations of the predictor of interest, $\beta\in \mathbb{R}^p$, $X$ is an $n\times p$ matrix of observations of control variables that are selected by the LASSO objective function and $u$ is an $n\times 1$ vector of independent and identically distributed error terms that is independent of $X$.  With this knowledge of the DGP, the target parameter of inferential inference after model selection can be written as $\mu_{T,n}(\widehat M_n)=\sqrt{n}e_1^{\prime}(\mathbb{E}[W_{E,i}W_{E,i}^{\prime}])^{-1}\mathbb{E}[W_{E,i}W_i^{\prime}]\delta$ for $E=\hat E_n$, where $W=(z,X)$ and $\delta=(\theta,\beta^{\prime})^{\prime}$.

For this simulation study, I generate data that entail significant departures from Gaussianity in a relatively small sample of $n=50$ in order to assess the relevance of the asymptotic guarantees provided by Proposition \ref{prop: uncond coverage} and the relative performance of the HySI CI in a small sample setting.  Across all simulation designs, I set $\alpha=0.05$ for nominal CI coverage of 95\%, $p=10$ potential control variables, $\delta=0$ and $\beta=(-4,4,0\ldots,0)^{\prime}$.  All quantities are computed across $1,000$ independent simulation replications.  The full matrix of regressors $W$ is generated in two ways.  In the ``independent'' case, the columns of $W$ are drawn from independent distributions, where each column is drawn from an independent $\mathcal{N}(0,1)$, Bernoulli or skew normal $(0,1,5)$ distribution with equal probability.  Each column is then normalized to have a sample mean of zero and unit Euclidean norm.  In the ``dependent'' case, each row of $W$ is generated independently from a multivariate normal distribution with mean vector zero and covariance matrix $(e^{-0.1|i-j|})_{1\leq i,j\leq p}$.  Each column is subsequently normalized to have unit Euclidean norm.  In each of these two cases, $Y$ is then generated according to \eqref{reg eqn} after sampling the entries of $u$ independently in four ways:~from a normal, skew normal (with shape parameter five), Laplace or uniform distribution, all with mean zero and unit variance.  

In each simulated data set, I perform the LASSO model selection exercise described in Section \ref{sec:LASSO} for several different values of the LASSO penalty parameter $\lambda\in\{1,2,4,8,16,100\}$ and construct CIs for the target parameter $\mu_{T,n}(\widehat M_n)$ ($\mu_{T,n}(\widehat M_{n/2})$ for the split-sample CI, see below) selected by the LASSO objective function.  Specifically, I calculate the naive CI that ignores model selection, a split-sample CI, the selective CI, the HySI CI using the recommended value of $\gamma=\alpha/10$ and the PoSI CI.  The naive CI is simply based on inverting the standard asymptotic $t$-test at the nominal level.  In accordance with the previous section, both the HySI and PoSI CIs computed in these simulations use the less conservative PoSI construction that incorporates the fact that the regressor $z$ is the predictor of interest and therefore not subject to selection.  The split-sample CI is constructed as follows:~the first $n/2$ observations are used to select the model, yielding $\widehat M_{n/2}$, while the final $n/2$ observations are used to construct a standard CI based on inverting the standard asymptotic $t$-test at the nominal level.  It is important to note that although the split-sample CI is known to have correct asymptotic coverage, it is not for the same object of interest, $\mu_{T,n}(\widehat M_n)$, for which the other CIs are designed.  Instead, the spit-sample CI has correct asymptotic coverage for the scaled population regression coefficient evaluated at the model selected by the first half of the data only, i.e.,~$\mu_{T,n}(\widehat M_{n/2})$.  This is especially important to keep in mind when evaluating the tradeoffs of the various CIs because selecting the model from only a portion of the data will yield a selected model with less desirable statistical properties (e.g.,~larger prediction errors).  Nevertheless, we include this comparison since it is commonly used as a valid method for inference after model selection.

The results of the Monte Carlo experiments are very similar across some of the error distributions and LASSO penalty parameters.  To save on space, I report a subset of results that illustrate the main features and tradeoffs of the full set of experiments.  In particular, I report results for the normal and skew normal error distributions and $\lambda\in\{1,4,16\}$.  

To begin, Table \ref{table:cov probs} displays the simulated (unconditional) coverage probabilities of the five CIs for the three different penalty parameter values.  The selective, HySI and PoSI CIs all have finite-sample coverage close to the nominal level of 95\%, where the selective and HySI CIs tend to slightly under-cover and the PoSI CIs tend to slightly over-cover.  The small coverage distortions of the selective and HySI CIs are to be expected from such small non-Gaussian data sets and they diminish for larger samples.  On the other hand, both the naive and split-sample CIs exhibit more sizable under-coverage.  This under-coverage is to be expected from the naive CI since it is known to incorrectly cover after using the data to select the model, even in large samples.  The split-sample CIs only exhibit notable coverage distortions in the dependent design cases.  Since split-sample CIs are known to have correct coverage in large samples, this is likely due to the fact that these CIs are constructed using only $n/2=25$ data points and the strong positive correlation between the regressors effectively reduces this sample size further relative to the independent design cases.

\begin{figure}[p]
\centering
\includegraphics[scale=0.27]{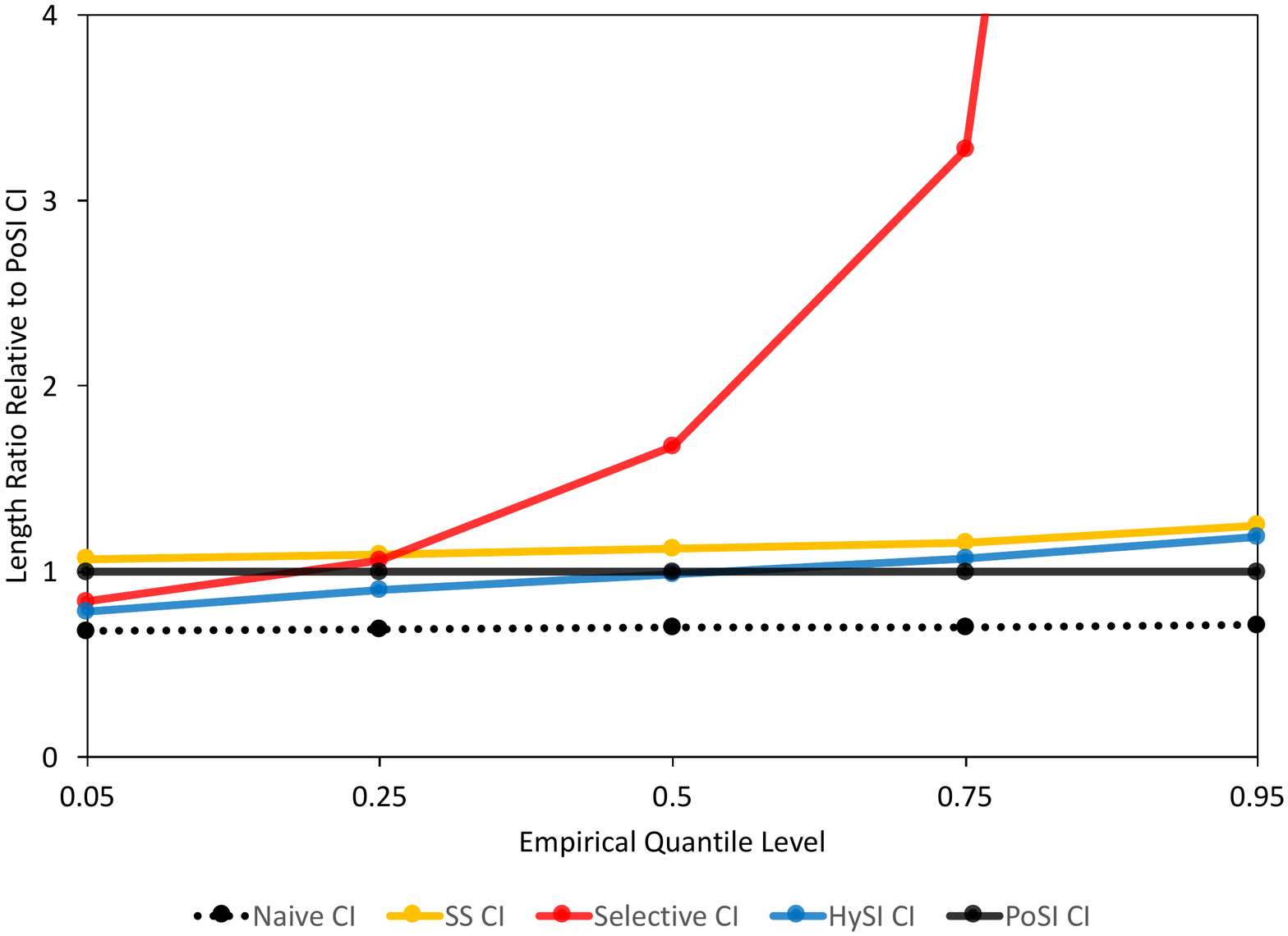}
\includegraphics[scale=0.27]{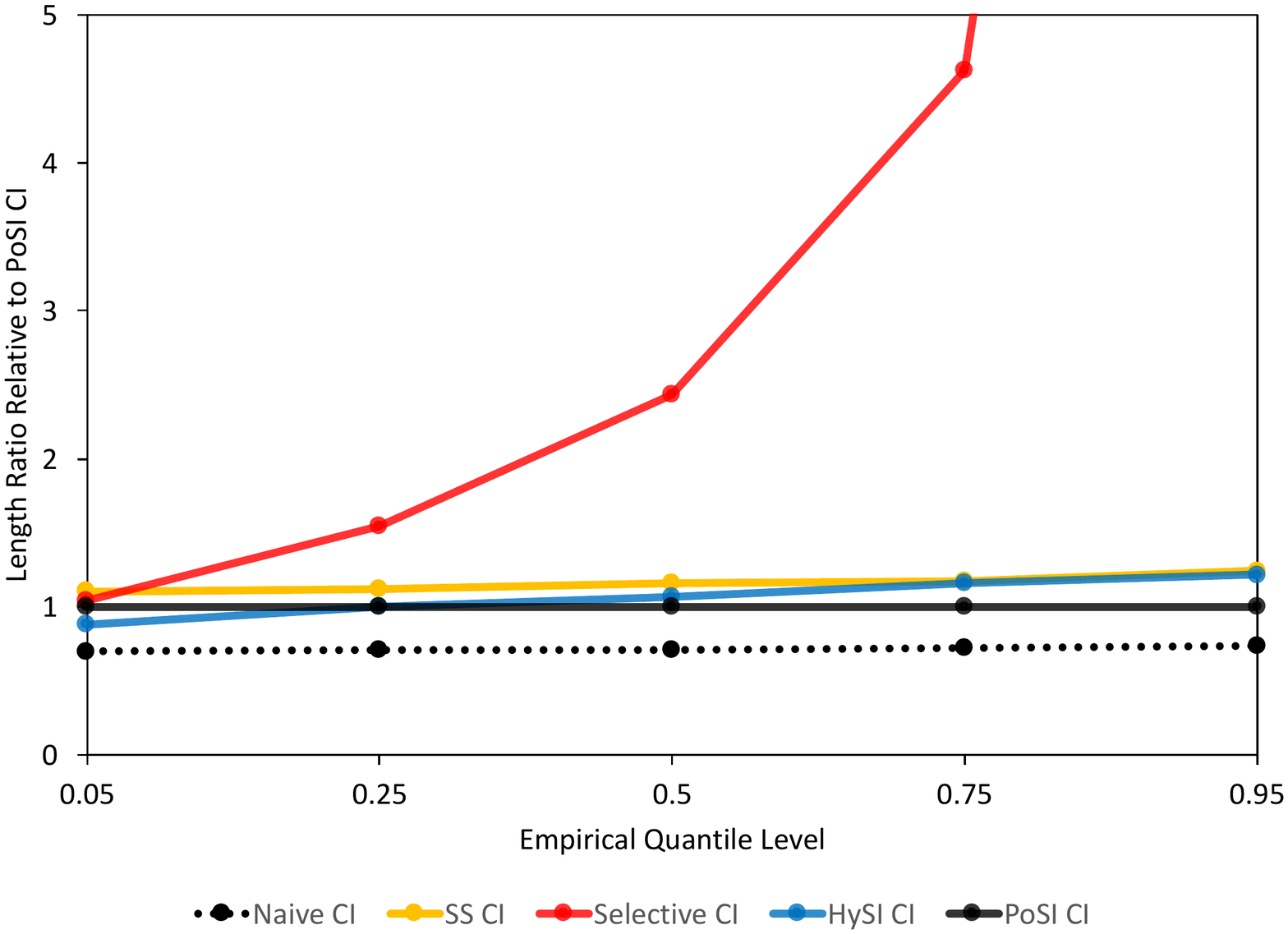}
\includegraphics[scale=0.27]{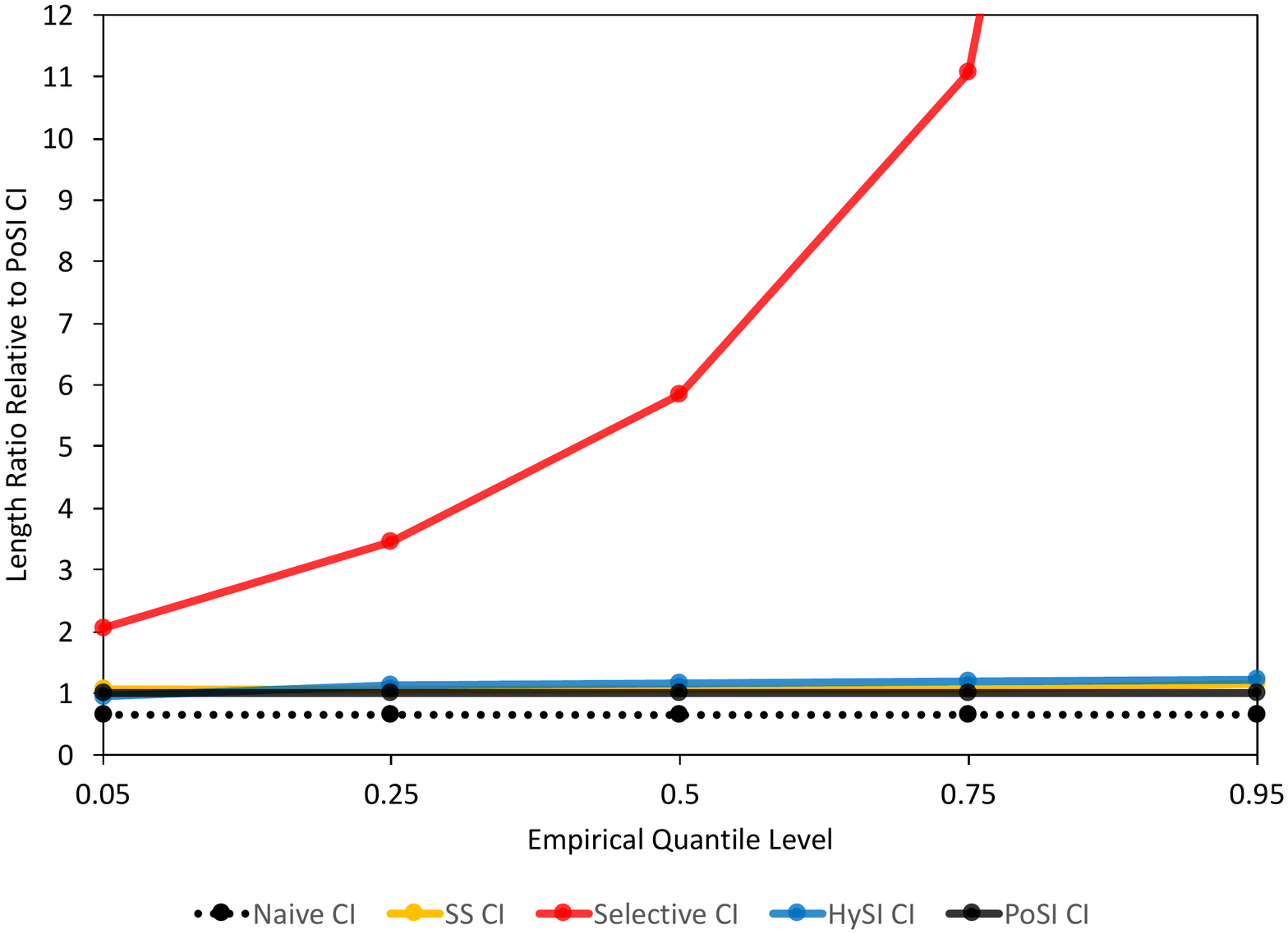}
\includegraphics[scale=0.27]{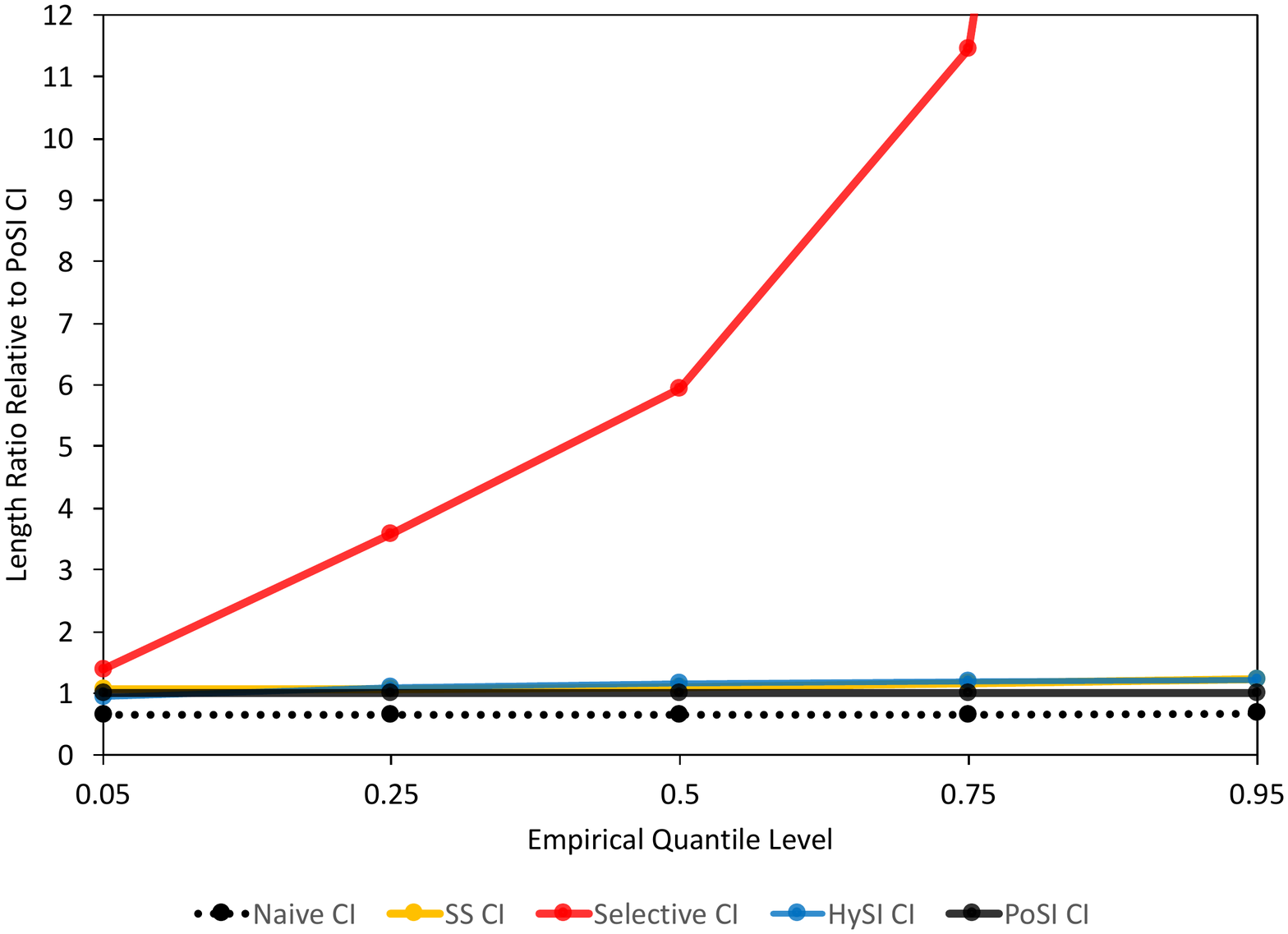}
\caption{\footnotesize This figure plots the $5^{th}$, $25^{th}$, $50^{th}$, $75^{th}$ and $95^{th}$ empirical quantiles of the lengths of the 95\% naive (dotted black), split-sample (orange), selective (red), HySI (blue) and PoSI (black) CIs divided by the corresponding length quantiles of the 95\% PoSI CI across Monte Carlo replications for inference after using LASSO to choose the control variables in the regression with a sample size of $n=50$ and the LASSO penalization parameter set to $\lambda=1$.  The \textbf{upper-left} plot corresponds to a design matrix with independent columns and error terms with a normal distribution.  The \textbf{upper-right} plot corresponds to a design matrix with independent columns and error terms with a skew normal distribution.  The \textbf{lower-left} plot corresponds to a design matrix with correlated columns and error terms with a normal distribution.  The \textbf{lower-right} plot corresponds to a design matrix with correlated columns and error terms with a skew normal distribution.\label{fig:lam=1}}
\end{figure}

\begin{figure}[p]
\centering
\includegraphics[scale=0.27]{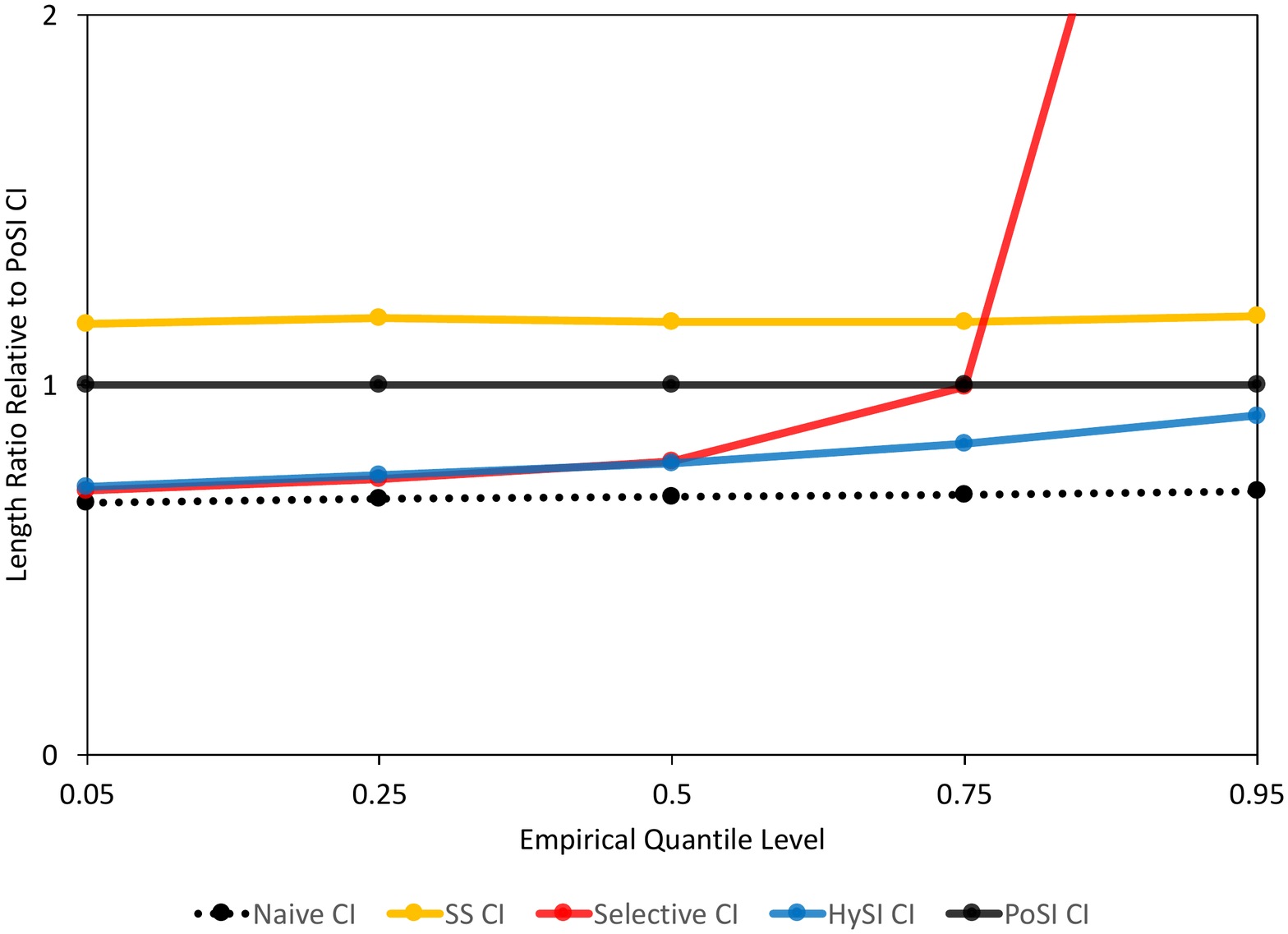}
\includegraphics[scale=0.27]{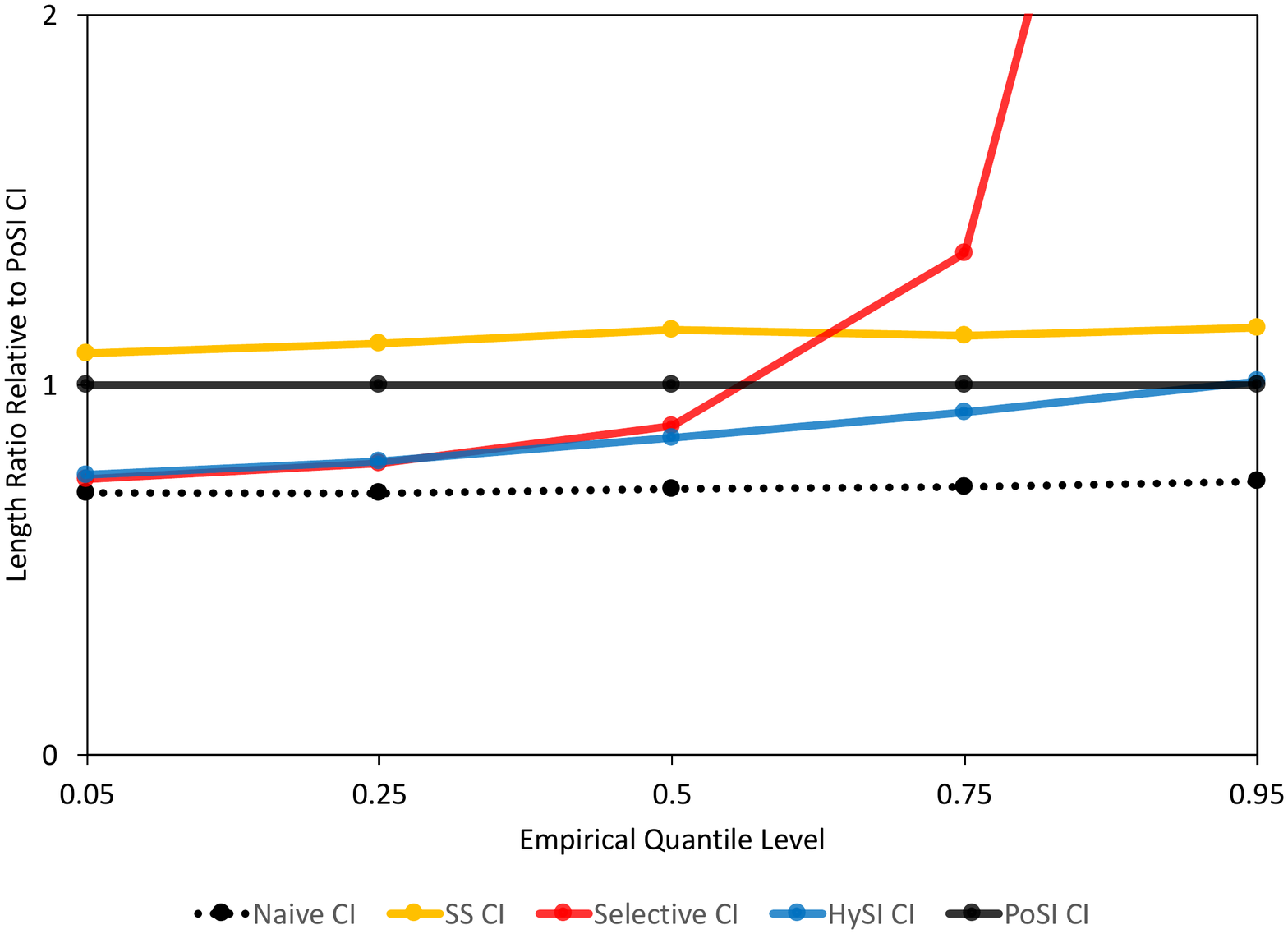}
\includegraphics[scale=0.27]{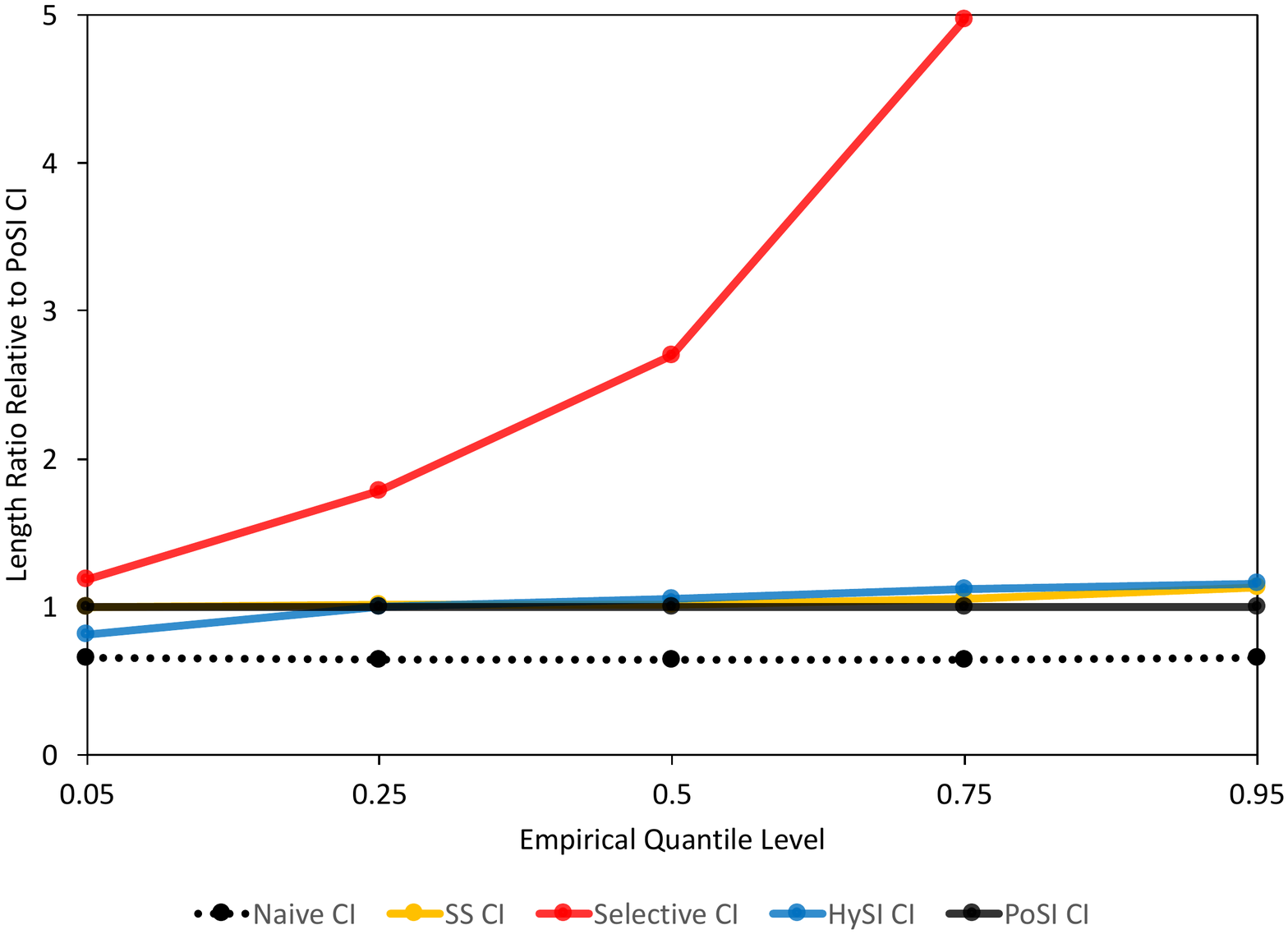}
\includegraphics[scale=0.27]{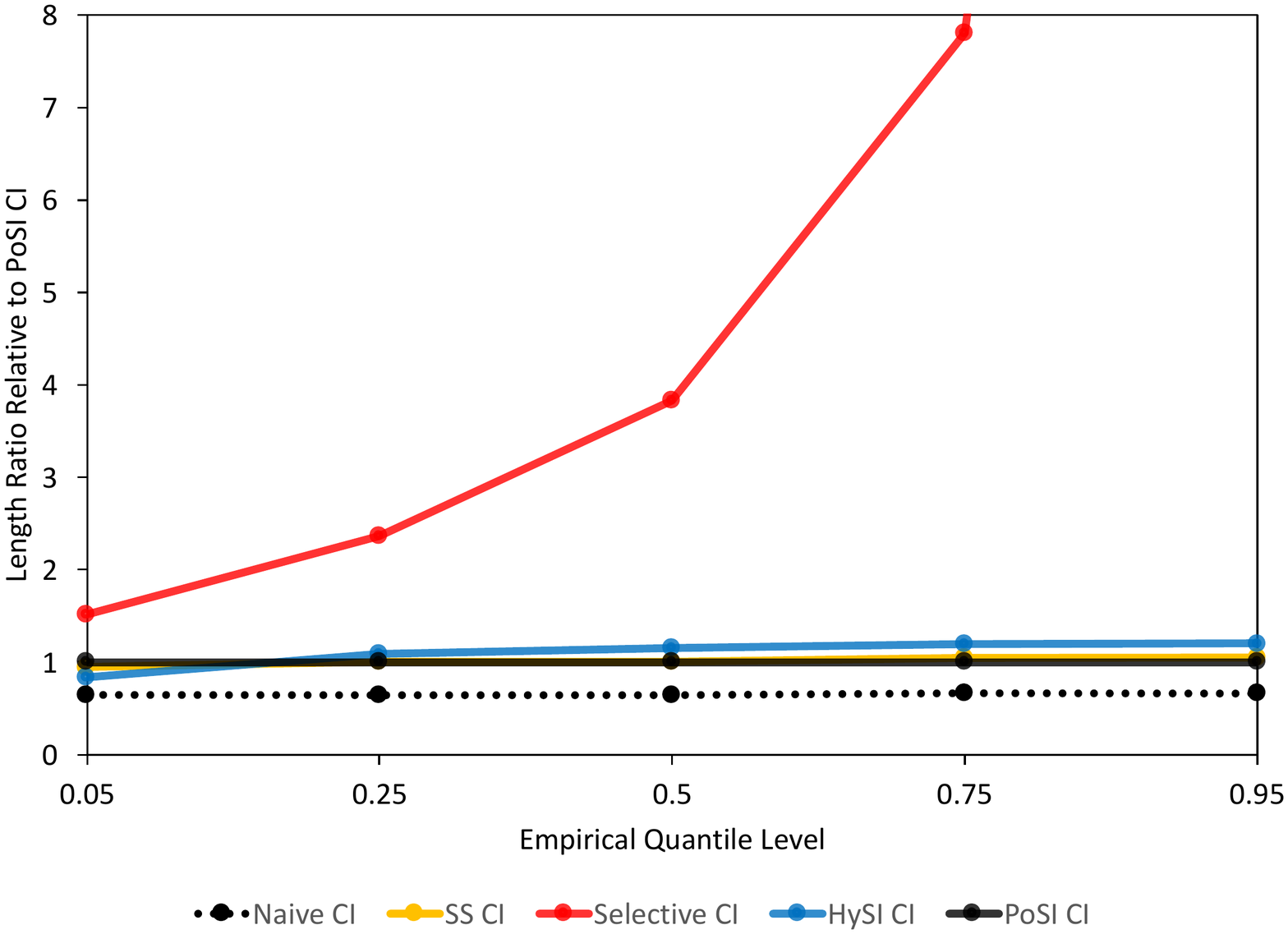}
\caption{\footnotesize This figure plots the $5^{th}$, $25^{th}$, $50^{th}$, $75^{th}$ and $95^{th}$ empirical quantiles of the lengths of the 95\% naive (dotted black), split-sample (orange), selective (red), HySI (blue) and PoSI (black) CIs divided by the corresponding length quantiles of the 95\% PoSI CI across Monte Carlo replications for inference after using LASSO to choose the control variables in the regression with a sample size of $n=50$ and the LASSO penalization parameter set to $\lambda=4$.  The \textbf{upper-left} plot corresponds to a design matrix with independent columns and error terms with a normal distribution.  The \textbf{upper-right} plot corresponds to a design matrix with independent columns and error terms with a skew normal distribution.  The \textbf{lower-left} plot corresponds to a design matrix with correlated columns and error terms with a normal distribution.  The \textbf{lower-right} plot corresponds to a design matrix with correlated columns and error terms with a skew normal distribution.\label{fig:lam=4}}
\end{figure}

\begin{figure}[p]
\centering
\includegraphics[scale=0.27]{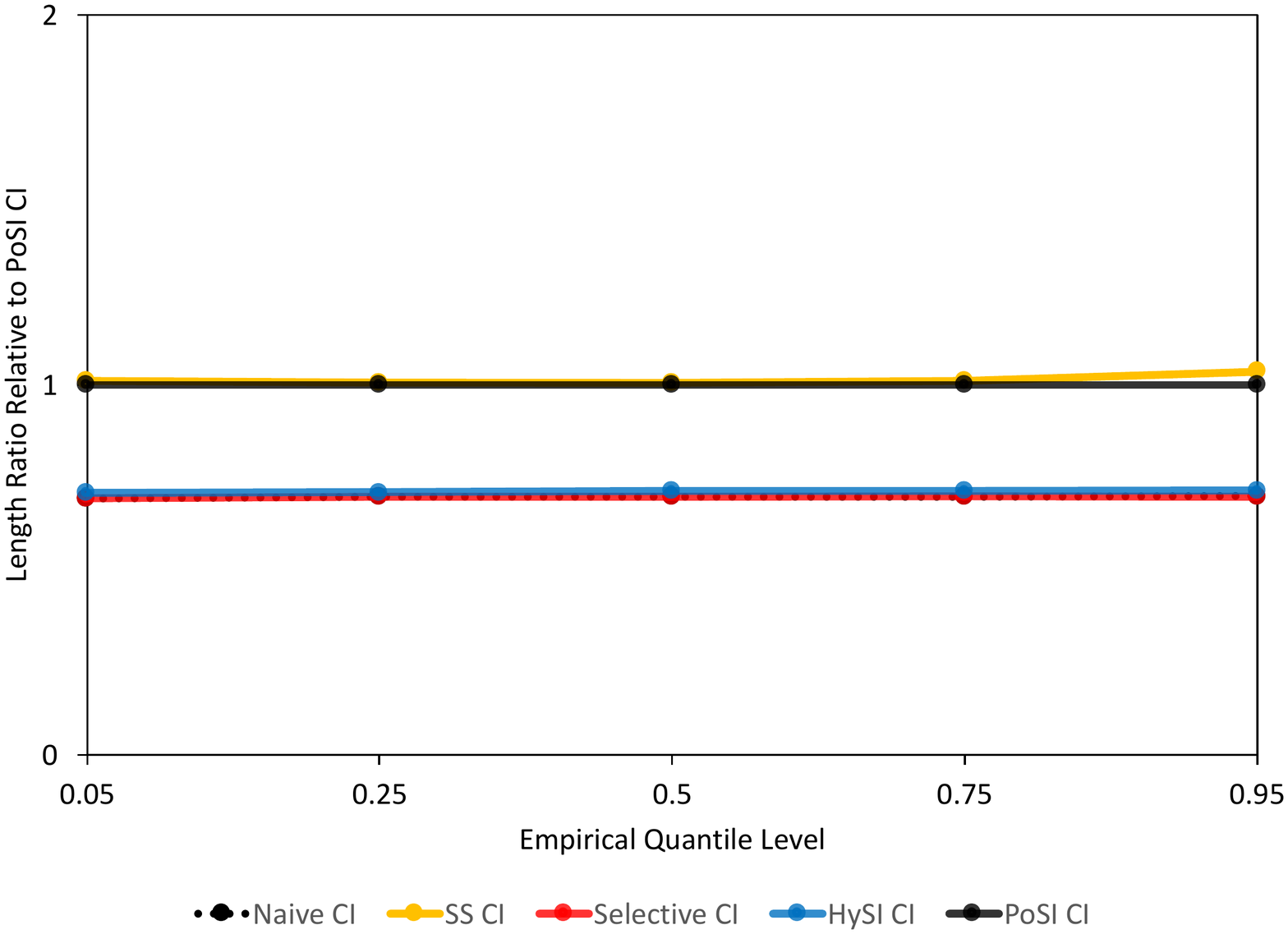}
\includegraphics[scale=0.27]{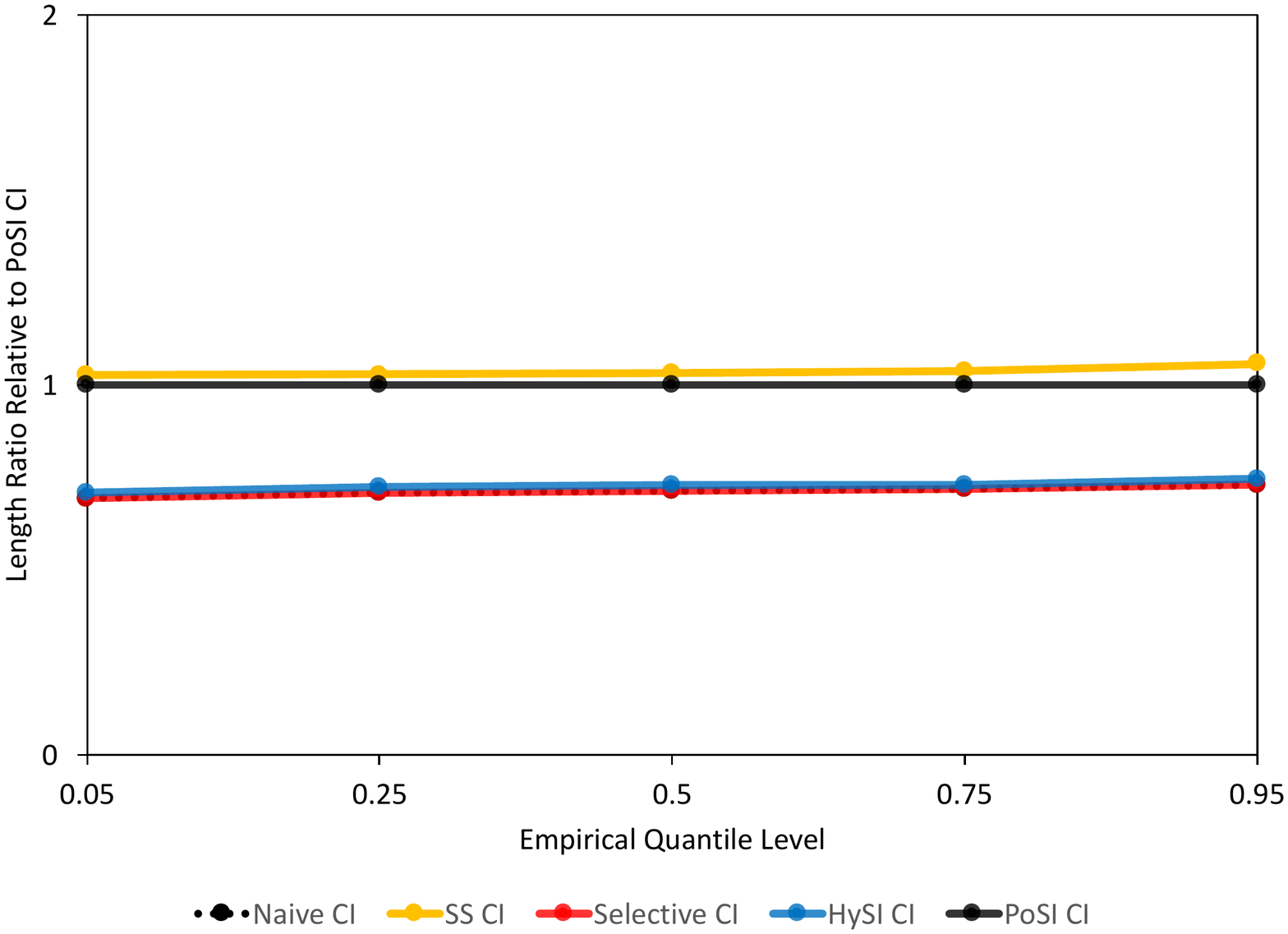}
\includegraphics[scale=0.27]{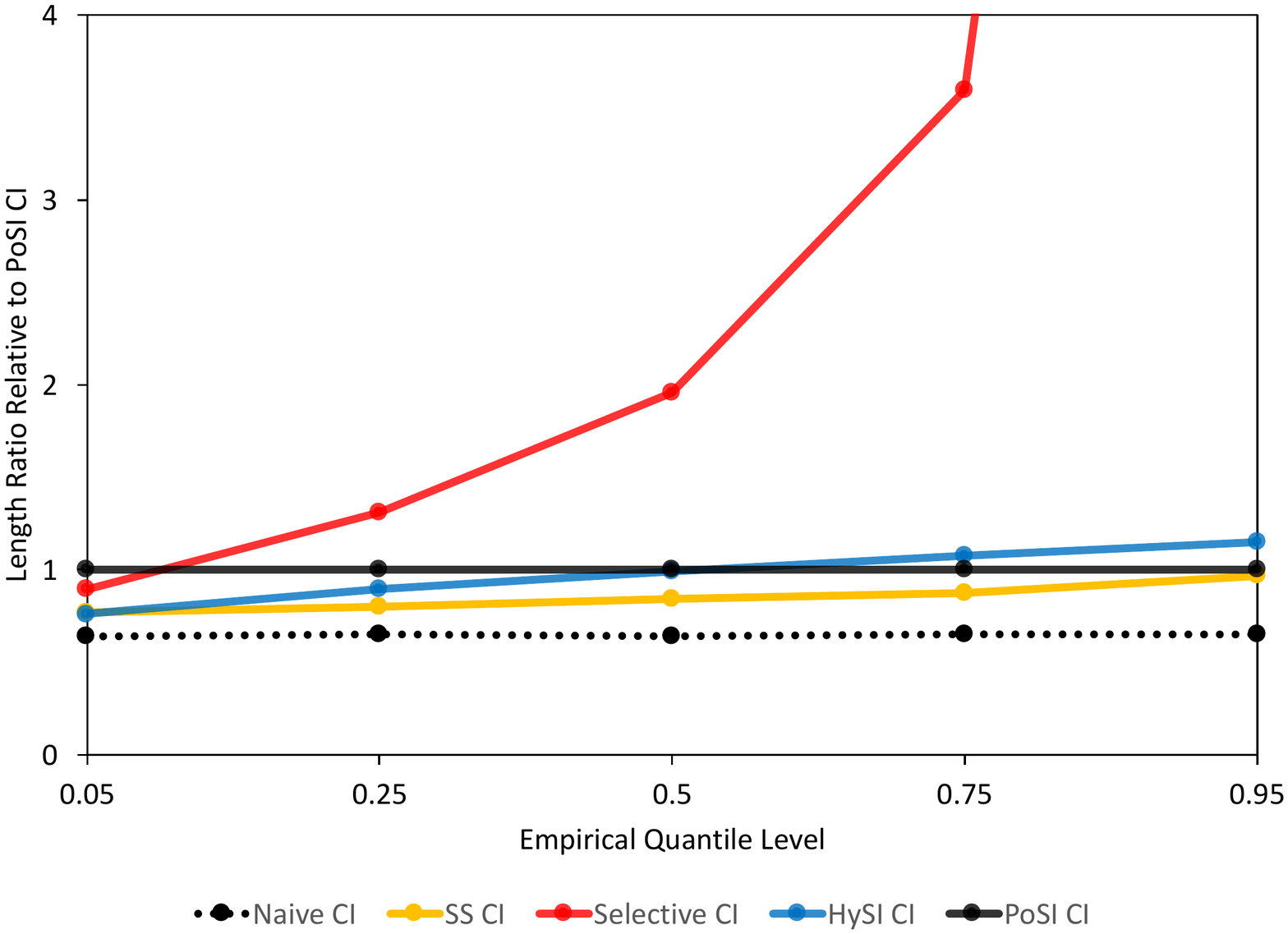}
\includegraphics[scale=0.27]{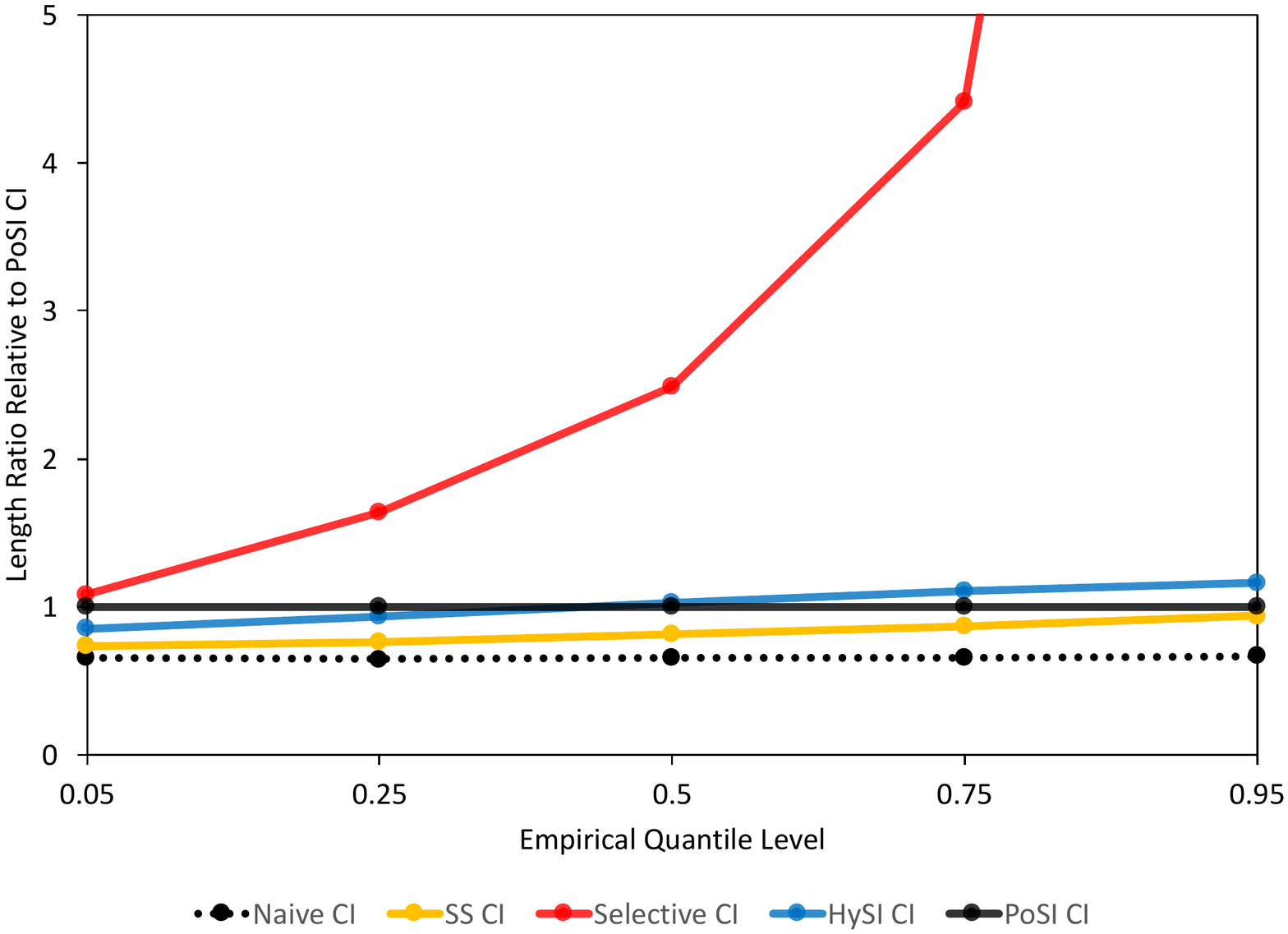}
\caption{\footnotesize This figure plots the $5^{th}$, $25^{th}$, $50^{th}$, $75^{th}$ and $95^{th}$ empirical quantiles of the lengths of the 95\% naive (dotted black), split-sample (orange), selective (red), HySI (blue) and PoSI (black) CIs divided by the corresponding length quantiles of the 95\% PoSI CI across Monte Carlo replications for inference after using LASSO to choose the control variables in the regression with a sample size of $n=50$ and the LASSO penalization parameter set to $\lambda=16$.  The \textbf{upper-left} plot corresponds to a design matrix with independent columns and error terms with a normal distribution.  The \textbf{upper-right} plot corresponds to a design matrix with independent columns and error terms with a skew normal distribution.  The \textbf{lower-left} plot corresponds to a design matrix with correlated columns and error terms with a normal distribution.  The \textbf{lower-right} plot corresponds to a design matrix with correlated columns and error terms with a skew normal distribution.\label{fig:lam=16}}
\end{figure}

Next, Figures \ref{fig:lam=1}--\ref{fig:lam=16} plot the ratios of the $5^{th}$, $25^{th}$, $50^{th}$, $75^{th}$ and $95^{th}$ empirical quantiles across simulation draws of the lengths of the five CIs relative to those corresponding to the PoSI CI for $\lambda=1$ (Figure \ref{fig:lam=1}), $\lambda=4$ (Figure \ref{fig:lam=4}) and $\lambda=16$ (Figure \ref{fig:lam=16}).  There are four panels within each figure corresponding to how the design matrix and errors are generated.  The ratios of the length quantiles of the PoSI CI relative to itself, which is always equal to one, is also included in the figures in order to detect when the other CIs' length quantiles are shorter than those of the PoSI CI.  Even though the naive CIs do not have correct coverage in large samples, I also include their ratios of length quantiles as a lower bound to show how close the other CIs come to attaining it.

From Figure \ref{fig:lam=1} we can see that for a small LASSO penalty parameter, the length quantiles of the selective CI are uniformly dominated by all other CIs.  For this low level of penalization the probability that LASSO chooses any given model is low, leading to excessively long selective CIs.  On the other hand, the HySI CIs tend to have similar length properties to those of the PoSI CIs, with some small increases in the dependent design cases (lower two panels).  The split-sample CIs also have similar length properties although, unlike the HySI CIs, their length quantiles always exceed those of the PoSI CIs with the additional drawback that their target model of interest is much less precisely selected.

Figure \ref{fig:lam=4} displays somewhat similar features to Figure \ref{fig:lam=1} although for the moderate-sized penalty parameter corresponding to this figure, we can start to see that the HySI CIs become notably shorter than the PoSI CIs when the selective CIs are shorter.  In the top two panels corresponding to an independent design matrix, the selective CIs tend to be shorter than in Figure \ref{fig:lam=1} because the higher penalty parameter increases the probability that LASSO chooses a given model.  In combination with the results in Figure \ref{fig:lam=1}, we can start to see the benefits of hybridization:~the HySI CI borrows the strengths of both the selective and PoSI CIs for different realizations of the data.  On the other hand, the bottom two panels display similar features to those in Figure \ref{fig:lam=1} because the dependent design matrices effectively reduce the signal-to-noise ratio in the model selection problem, making model selection probabilities significantly lower.

The top two panels of Figure \ref{fig:lam=16} clearly show the benefits of using HySI instead of PoSI when model selection probabilities are high (due to the large penalty parameter).  Here we can see that the length quantiles of both the selective and HySI CIs are nearly identical to those of the naive CI and substantially smaller than those of the PoSI CI.  This is a clear illustration that the HySI CIs attain nearly the same short lengths as the selective CI when they are short while guarding against the excessive lengths of the latter for unfavorable realizations of the data.  From the bottom two panels of the figure, we can see that the selective CIs still tend to be significantly longer and the HySI CIs tend to be similar to the PoSI CIs with a dependent design matrix.  This is again due to a reduced signal-to-noise ratio relative to the size of the penalty parameter.  When the penalty parameter is increased to $\lambda=100$, the relative length quantiles for the dependent design matrices look very similar to those in the top two panels of this figure with the length quantiles of the selective and HySI CIs being nearly indistinguishable from those of the naive CIs, entailing length reductions of 34--35\% relative to the PoSI CI across all quantile levels.


\section{Empirical Application to Diabetes Data}\label{sec:empirical}

I further investigate the properties of HySI CIs in an empirical application to the diabetes data set from \cite{EHJT04}.  This data set was also examined by \cite{LSST16} in their empirical application of selective inference after using LASSO as a model selection device, thus serving as a benchmark application for inference after using LASSO.  Departing from the exact exercise performed by \cite{LSST16}, I perform the LASSO model selection exercise described in Section \ref{sec:LASSO} multiple times for the response of interest $y$ being equal to a quantitative measure of disease progression one year after baseline.  In each empirical exercise, I set one of the 10 regressors in the data set as a predictor of interest $z$ while allowing the remaining nine regressors to be potential control variables $X$ selected by LASSO.  I perform these exercises for two values of the LASSO penalty parameter $\lambda\in\{50,190\}$ to illustrate the merits of the HySI CIs relative to other CIs under ``low'' and ``high'' levels of penalization.  The penalty parameter of $\lambda=190$ was examined by \cite{LSST16} for this data set and corresponds to LASSO selecting three to four control variables across the different predictors of interest.  On the other hand, $\lambda=50$ corresponds to LASSO selecting six to seven controls.

Figures \ref{fig:app,lam=50} and \ref{fig:app,lam=190} plot the naive, split-sample (using half of the data for model selection), selective, HySI (with $\gamma=\alpha/10$) and PoSI nominal 95\% CIs for $\lambda=50$ and $\lambda=190$ and each of the 10 predictors of interest:~Age, Sex, Body-Mass Index (BMI), Blood Pressure (BP) and six different blood serum measurements (S1--S6).  Before comparing the CIs, I reiterate that the naive CI does not have correct 95\% coverage and that the split-sample CIs cover a different, arguably inferior, target of interest based upon a model selected with half as much data.  Figure \ref{fig:app,lam=50} provides a striking illustration of how much shorter the HySI CIs can become relative to selective CIs at this lower level of LASSO penalization:~the HySI CI is shorter than the selective CI across all predictors of interest, with a length averaging 52\% of the latter across the predictors and several length reductions in excess of 65\%.  In comparison to the PoSI CIs, the HySI CIs tend to be very similar across predictors of interest with an average length increase of 2\%.

In contrast to Figure \ref{fig:app,lam=50}, Figure \ref{fig:app,lam=190} illustrates a more favorable case for the selective CIs at this higher level of LASSO penalization.  For all but one predictor of interest (S5), the HySI CIs are very similar to the selective CIs in these cases where it performs well, entailing slight length increases over the latter of 0--3\%.  However, the selective CI for S5 is excessively long while the HySI CI for this same predictor is not, providing a length reduction of nearly 65\%.  In comparison to the PoSI CIs, the HySI CIs are shorter for all predictors of interest with an average length reduction in excess of 25\% across predictors and reaching more than 35\% for several of them.

In summary, Figures \ref{fig:app,lam=50}--\ref{fig:app,lam=190} provide real world evidence that HySI CIs perform very similarly to selective CIs (and also naive CIs) in scenarios that are favorable to the latter while transitioning more closely to PoSI CIs in scenarios for which selective CIs become very long.

\section{Conclusions and Extensions}\label{sec:conclusion}

In this paper, I introduce an alternative CI for inference after model selection to those that currently dominate the literature.  By relaxing the coverage requirement of selective CIs, these HySI CIs are able to borrow upon the relative strengths of selective and PoSI CIs to yield CIs with desirable length properties across a wide variety of data realizations.

Two questions that I did not address in this paper but may be worth investigating in follow-up research are whether PoSI CIs that do not satisfy the structure imposed by Assumption \ref{ass: PoSI CS} can be used as an ingredient in the construction of HySI CIs and whether HySI CIs can be constructed to have correct asymptotic coverage for high-dimensional models with a diverging number of parameters.  The first question is interesting in light of recent work dedicated to producing PoSI CIs that are either shorter and/or easier to compute in the presence of many models under consideration (see e.g.,~\citealp{KBBCGZ19}).  For the second question, results in \cite{TRTW18} suggest that uniform asymptotic coverage of HySI CIs may not be possible in high-dimensional models.  On the other hand, results in \cite{TT17} suggest that point-wise asymptotic coverage may be attainable.

\newpage

\bibliographystyle{apalike}
\bibliography{hybrid_post_sel}


\newpage

\begin{table}[htbp]
\caption{Unconditional Coverage Probabilities}\label{table:cov probs}
\centering{}%
\begin{tabular}{lcccccc}
 & & \multicolumn{5}{c}{Confidence Interval}\tabularnewline
 $\lambda$ &  & Naive & SS  & Sel & HySI & PoSI \tabularnewline
\hline
\multicolumn{7}{c}{Indep Design, Normal Errors}  \tabularnewline
1  &  &  0.88	&0.93	&0.92	&0.92	&0.98  \tabularnewline
4  &  & 0.91	&0.94	&0.91	&0.91	&0.98  \tabularnewline
16  &  &  0.93	&0.95	&0.93	&0.93	&0.99  \tabularnewline
\multicolumn{7}{c}{Indep Design, Skew Normal Errors} \tabularnewline
1  &  &  0.90	&0.93	&0.94	&0.94	&0.97  \tabularnewline
4  &  &  0.92	&0.95	&0.93	&0.93	&0.99  \tabularnewline
16  &  &  0.92	&0.94	&0.92	&0.93	&0.99  \tabularnewline
\multicolumn{7}{c}{Dep Design, Normal Errors}  \tabularnewline
1  &  &  0.87	&0.95	&0.96	&0.96	&0.99  \tabularnewline
4  &  & 0.89	&0.91	&0.93	&0.93	&0.98  \tabularnewline
16  &  &  0.90	&0.85	&0.92	&0.92	&0.99  \tabularnewline
\multicolumn{7}{c}{Dep Design, Skew Normal Errors} \tabularnewline
1  &  &  0.89	&0.93	&0.95	&0.95	&0.98  \tabularnewline
4  &  &  0.88	&0.92	&0.94	&0.94	&0.98  \tabularnewline
16  &  &  0.92	&0.89	&0.93	&0.93	&0.99  \tabularnewline
\hline
\end{tabular}
\caption*{\footnotesize This table reports unconditional coverage probabilities for the selected population coefficient on the predictor of interest after using LASSO to choose the control variables in the regression across Monte Carlo replications, all evaluated at the nominal coverage level of 95\%.  Coverage probabilities are reported for naive, split-sample (``SS''), selective (``Sel''), HySI and PoSI confidence intervals for a sample size of $n=50$.  The coverage probabilities are reported for three different values of the LASSO penalization parameter $\lambda\in\{1,4,16\}$.  The design matrix is generated with independent (upper half of table) or correlated (lower half) columns and the error terms have normal or skew normal distributions.}
\end{table}

\begin{figure}[p]
\centering
\includegraphics[scale=0.65]{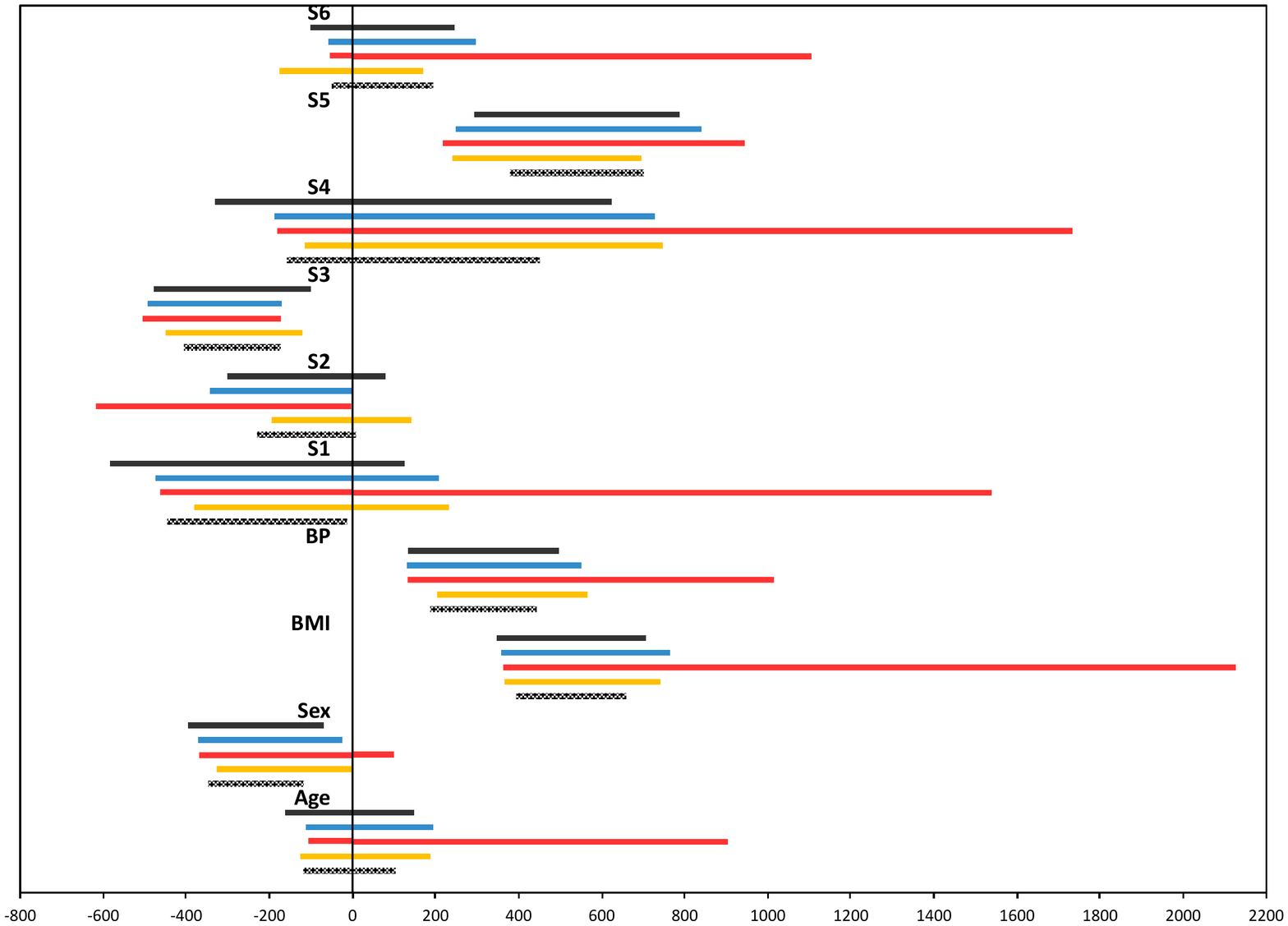}
\caption{\footnotesize This figure plots the naive (dotted black), split-sample (orange), selective (red), HySI (blue) and PoSI (black) CIs for the population coefficient for each of the 10 regressors in the diabetes data set after using LASSO to choose the control variables in the regression with the LASSO penalization parameter set to $\lambda=50$.  \label{fig:app,lam=50}}
\end{figure}

\begin{figure}[p]
\centering
\includegraphics[scale=0.65]{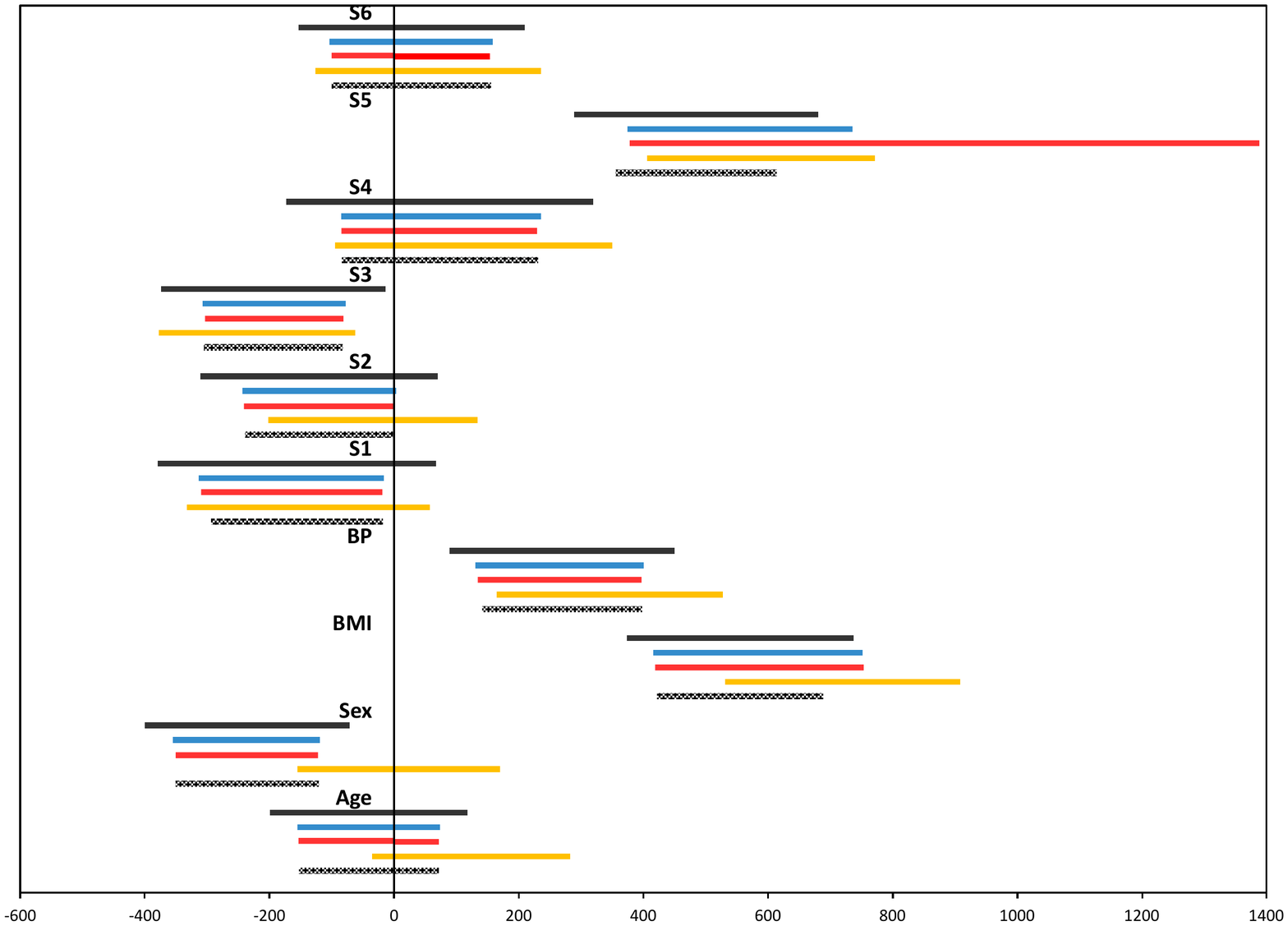}
\caption{\footnotesize This figure plots the naive (dotted black), split-sample (orange), selective (red), HySI (blue) and PoSI (black) CIs for the population coefficient for each of the 10 regressors in the diabetes data set after using LASSO to choose the control variables in the regression with the LASSO penalization parameter set to $\lambda=190$.  \label{fig:app,lam=190}}
\end{figure}

\end{document}


\def\spacingset#1{\renewcommand{\baselinestretch}%
{#1}\small\normalsize} \spacingset{1}

\title{Supplemental Appendix for ``Hybrid Confidence Intervals for Informative Uniform Asymptotic Inference After Model Selection''} 


\author{Adam McCloskey\footnote{Department of Economics, University of Colorado, adam.mccloskey@colorado.edu}
}

\maketitle

\begin{abstract}
This supplemental appendix contains the proofs of the theoretical results in ``Hybrid Confidence Intervals for Informative Uniform Asymptotic Inference After Model Selection.''
\end{abstract}

\textbf{Proof of Lemma 1:} By Assumption 1,
\begin{align*}
\left\{\widehat{M}_n=M\right\}&=\{A_MD_n(M)\leq \widehat a_{M,n}\} \\
&=\left\{A_M\left(\widehat \Sigma_{DT,n}^{(M)}/\widehat\Sigma_{T,n}(M,M)\right)T_n(M)\leq \widehat a_{M,n}-A_MZ_{M,n}\right\} \\
&=\left\{\left(A_M\left(\widehat \Sigma_{DT,n}^{(M)}/\widehat\Sigma_{T,n}(M,M)\right)\right)_jT_n(M)\leq \widehat a_{M,n,j}-(A_MZ_{M,n})_j\right\} \\
&=\left\{\begin{array}{ll}
T_n(M)\leq \frac{\widehat a_{M,n,j}-(A_MZ_{M,n})_j}{(A_M\widehat \Sigma_{DT,n}^{(M)}/\widehat\Sigma_{T,n}(M,M))_j}, & \text{for } j:(A_M\widehat \Sigma_{DT,n}^{(M)}/\widehat\Sigma_{T,n}(M,M))_j>0 \\
T_n(M)\geq \frac{\widehat a_{M,n,j}-(A_MZ_{M,n})_j}{(A_M\widehat \Sigma_{DT,n}^{(M)}/\widehat\Sigma_{T,n}(M,M))_j}, & \text{for } j:(A_M\widehat \Sigma_{DT,n}^{(M)}/\widehat\Sigma_{T,n}(M,M))_j<0 \\
0\leq \widehat a_{M,n,j}-(A_MZ_{M,n})_j & \text{for } j:(A_M\widehat \Sigma_{DT,n}^{(M)}/\widehat\Sigma_{T,n}(M,M))_j=0
\end{array}\right\}.
\end{align*}
The statement of the lemma immediately follows. $\blacksquare$

The following lemma is useful for proving the correct uniform asymptotic coverage of the HySI CI $CI_{n,\widehat{M}_n}^{H,\alpha}$.

\setcounter{lemma}{1}

\begin{lemma} \label{lem: truncation recenter}
 For $Z_{M,n}^*=D_n(M)-\left(\widehat \Sigma_{DT,n}^{(M)}/\widehat\Sigma_{T,n}(M,M)\right)T_n^*(M)$ with $T_n^*(M)=T_n(M)-\mu_{T,n}(M)$, $\mathcal{V}_{M,n}^{-}(Z_{M,n}^*)=\mathcal{V}_{M,n}^{-}(Z_{M,n})-\mu_{T,n}(M)$ and $\mathcal{V}_{M,n}^{+}(Z_{M,n}^*)=\mathcal{V}_{M,n}^{+}(Z_{M,n})-\mu_{T,n}(M)$.
\end{lemma}

\textbf{Proof:} Noting that
\begin{align*}
Z_{M,n}^*&=D_n(M)-\left(\widehat \Sigma_{DT,n}^{(M)}/\widehat\Sigma_{T,n}(M,M)\right)T_n(M)+\left(\widehat \Sigma_{DT,n}^{(M)}/\widehat\Sigma_{T,n}(M,M)\right)\mu_{T,n}(M) \\
&=Z_{M,n}+\left(\widehat \Sigma_{DT,n}^{(M)}/\widehat\Sigma_{T,n}(M,M)\right)\mu_{T,n}(M),
\end{align*}
we have 
\begin{align*}
\mathcal{V}_{M,n}^{-}(Z_{M,n}^*)&=\max_{j:(A_M\widehat \Sigma_{DT,n}^{(M)}/\widehat\Sigma_{T,n}(M,M))_j<0}\frac{\widehat a_{M,n,j}-(A_MZ_{M,n}^*)_j}{(A_M\widehat \Sigma_{DT,n}^{(M)}/\widehat\Sigma_{T,n}(M,M))_j} \\
&=\max_{j:(A_M\widehat \Sigma_{DT,n}^{(M)}/\widehat\Sigma_{T,n}(M,M))_j<0}\frac{\widehat a_{M,n,j}-(A_MZ_{M,n})_j-\left(A_M\left(\widehat \Sigma_{DT,n}^{(M)}/\widehat\Sigma_{T,n}(M,M)\right)\mu_{T,n}(M)\right)_j}{(A_M\widehat \Sigma_{DT,n}^{(M)}/\widehat\Sigma_{T,n}(M,M))_j} \\
&=\max_{j:(A_M\widehat \Sigma_{DT,n}^{(M)}/\widehat\Sigma_{T,n}(M,M))_j<0}\frac{\widehat a_{M,n,j}-(A_MZ_{M,n})_j}{(A_M\widehat \Sigma_{DT,n}^{(M)}/\widehat\Sigma_{T,n}(M,M))_j}-\mu_{T,n}(M) \\
&=\mathcal{V}_{M,n}^{-}(Z_{M,n})-\mu_{T,n}(M).
\end{align*}
The proof for $\mathcal{V}_{M,n}^{+}(Z_{M,n}^*)$ is entirely analogous and therefore omitted. $\blacksquare$

\textbf{Proof of Proposition 1:} By the same argument used in the proof of Proposition 5 in \cite{AKM20}, 
\[F_{TN}(t;\mu,\Sigma_T(M,M),\mathcal{V}_{M,n}^{-,H}(z,\mu),\mathcal{V}_{M,n}^{+,H}(z,\mu))\] 
is decreasing in $\mu$ so that $\widehat\mu_{T,n}^{H,\frac{\alpha-\gamma}{2(1-\gamma)}}(\widehat{M}_n)\geq \mu_{T,n}(\widehat{M}_n)$ is equivalent to 
\begin{equation*}
F_{TN}\left(T_n(\widehat{M}_n);\mu_{T,n}(\widehat{M}_n),\widehat\Sigma_{T,n}(\widehat{M}_n,\widehat{M}_n),\mathcal{V}_{\widehat M_n,n}^{-,H}(Z_{\widehat M_n,n},\mu_{T,n}(\widehat{M}_n)),\mathcal{V}_{\widehat M_n,n}^{+,H}(Z_{\widehat M_n,n},\mu_{T,n}(\widehat{M}_n))\right)\geq 1-\frac{\alpha-\gamma}{2(1-\gamma)}. 
\end{equation*}
Further, Lemma \ref{lem: truncation recenter} implies
\begin{align*}
& F_{TN}\left(T_n(\widehat{M}_n);\mu_{T,n}(\widehat{M}_n),\widehat\Sigma_{T,n}(\widehat{M}_n,\widehat{M}_n),\mathcal{V}_{\widehat M_n,n}^{-,H}(Z_{\widehat M_n,n},\mu_{T,n}(\widehat{M}_n)),\mathcal{V}_{\widehat M_n,n}^{+,H}(Z_{\widehat M_n,n},\mu_{T,n}(\widehat{M}_n))\right) \\
& =F_{TN}\left(T_n^*(\widehat{M}_n)+\mu_{T,n}(\widehat{M}_n);\mu_{T,n}(\widehat{M}_n),\widehat\Sigma_{T,n}(\widehat{M}_n,\widehat{M}_n),\right. \\
&\qquad\max\left\{\mathcal{V}_{\widehat M_n,n}^{-}(Z_{\widehat M_n,n}^*)+\mu_{T,n}(\widehat{M}_n),\mu_{T,n}(\widehat{M}_n)-\sqrt{\widehat\Sigma_{T,n}(\widehat M_n,\widehat M_n)}K_{n,\gamma}\right\}, \\
&\qquad \left.\min\left\{\mathcal{V}_{\widehat M_n,n}^{+}(Z_{\widehat M_n,n}^*)+\mu_{T,n}(\widehat{M}_n),\mu_{T,n}(\widehat{M}_n)+\sqrt{\widehat\Sigma_{T,n}(\widehat M_n,\widehat M_n)}K_{n,\gamma}\right\}\right) \\
&=F_{TN}\left(T_n^*(\widehat{M}_n);0,\widehat\Sigma_{T,n}(\widehat{M}_n,\widehat{M}_n),\max\left\{\mathcal{V}_{\widehat M_n,n}^{-}(Z_{\widehat M_n,n}^*),-\sqrt{\widehat\Sigma_{T,n}(\widehat M_n,\widehat M_n)}K_{n,\gamma}\right\},\right. \\
&\qquad\left.\min\left\{\mathcal{V}_{\widehat M_n,n}^{+}(Z_{\widehat M_n,n}^*),\sqrt{\widehat\Sigma_{T,n}(\widehat M_n,\widehat M_n)}K_{n,\gamma}\right\}\right)
\end{align*}
so that $\widehat\mu_{T,n}^{H,\frac{\alpha-\gamma}{2(1-\gamma)}}(\widehat{M}_n)\geq \mu_{T,n}(\widehat{M}_n)$ is equivalent to
\begin{align}
&F_{TN}\left(T_n^*(\widehat{M}_n);0,\widehat\Sigma_{T,n}(\widehat{M}_n,\widehat{M}_n),\max\left\{\mathcal{V}_{\widehat M_n,n}^{-}(Z_{\widehat M_n,n}^*),-\sqrt{\widehat\Sigma_{T,n}(\widehat M_n,\widehat M_n)}K_{n,\gamma}\right\},\right. \notag \\
&\qquad\left.\min\left\{\mathcal{V}_{\widehat M_n,n}^{+}(Z_{\widehat M_n,n}^*),\sqrt{\widehat\Sigma_{T,n}(\widehat M_n,\widehat M_n)}K_{n,\gamma}\right\}\right)\geq 1-\frac{\alpha-\gamma}{2(1-\gamma)}. \label{eq: trunc normal equiv}
\end{align}

By an extension of Lemma 5 of \cite{AKM20}, to prove the statment of the proposition, it suffices to show that for all subsequences $\{n_s\}\subset \{n\}$, $\{\mathbb{P}_{n_s}\}\in \times_{n=1}^{\infty}\mathcal{P}_n$ with
\begin{enumerate}
\item $\Sigma(\mathbb{P}_{n_s})\rightarrow \Sigma^*\in\mathcal{S}$
\begin{align*}
\mathcal{S}&=\{\Sigma:1/\bar \lambda\leq \Sigma_T(M,M)\leq \bar\lambda,1/\bar \lambda\leq \lambda_{\min}(\Sigma_D^{(M)})\leq\lambda_{\max}(\Sigma_D^{(M)})\leq \bar\lambda\},
\end{align*}
\item $K_{\gamma}(\mathbb{P}_{n_s})\rightarrow K_{\gamma}^*\in [0,\bar\lambda]$,
\item $a_{M,n_s}(\mathbb{P}_{n_s})\rightarrow a_M^*\in [-\bar\lambda,\bar\lambda]^{\dim(a_M)}$,
\item $\mathbb{P}_{n_s}\left(\widehat{M}_{n_s}=M,\mu_{T,{n_s}}(\widehat{M}_{n_s})\in CI_{n_s,\widehat{M}_{n_s}}^{P,\gamma}\right)\rightarrow p^*\in(0,1]$, and
\item $\mu_{D,n_s}(M;\mathbb{P}_{n_s})\rightarrow \mu_D^*(M)\in [-\infty,\infty]^{\dim(D^*(M))}$
	\end{enumerate}
for some finite $\bar\lambda$, we have
\[\lim_{n\rightarrow\infty}\mathbb{P}_{n_s}\left(\mu_{T,n_s}(\widehat{M}_{n_s})\in CI_{n_s,\widehat{M}_{n_s}}^{\alpha,H}\left|\widehat{M}_{n_s}=M,\mu_{T,n_s}(\widehat{M}_{n_s})\in CI_{n_s,\widehat{M}_{n_s}}^{P,\gamma}\right.\right)=\frac{1-\alpha}{1-\gamma}.\]
Let $\{\mathbb{P}_{n_s}\}$ be a sequence satisfying conditions 1.--5.  Now under $\{\mathbb{P}_{n_s}\}$, $(T_{n_s}^*,\widehat{\Sigma}_{T,n_s},\widehat{\Sigma}_{DT,n_s},K_{n_s,\gamma})\overset{d}\longrightarrow (T^*,{\Sigma}_T^*,{\Sigma}_{DT}^*,K_{\gamma}^*)$ by Assumptions 2--4, where $T_n^*=(T_n^*(1),\ldots,T_n^*(|\mathcal{M}|)^{\prime}$ and $T^*\sim\mathcal{N}(0,\Sigma_T^*)$.  In addition, conditions 3.--4. along with Assumptions 1--2 imply that under $\{\mathbb{P}_{n_s}\}$, $\widehat M_{n_s}\overset{d}\longrightarrow \widehat M$, where $\widehat M=M\in \mathcal{M}$ if and only if $A_M(D^*(M)+\mu_D^*(M))\leq a_M^*$ with $D^*(M)\sim\mathcal{N}(0,\Sigma_D^{(M)*})$.  This convergence occurs jointly with that for $(T_{n_s}^*,\widehat{\Sigma}_{T,n_s},\widehat{\Sigma}_{DT,n_s},K_{n_s,\gamma})$.  Note that it is not possible for $(A_M\mu_D^*(M))_j=\infty$ for any $j$ under conditions 3.--4. and Assumptions 1--2.  Thus, under Assumptions 1--3, similar arguments to those used in the proof of Lemma 8 in \cite{AKM20} show that for any $M\in \mathcal{M}$, $(\mathcal{V}_{M,n_s}^{-}(Z_{M,n_s}^*),\mathcal{V}_{M,n_s}^{+}(Z_{M,n_s}^*))\overset{d}\longrightarrow (\mathcal{V}_{M}^{-}(Z_{M}^*),\mathcal{V}_{M}^{+}(Z_{M}^*))$ under $\{\mathbb{P}_{n_s}\}$, where $\mathcal{V}_{M}^{-}(z)$ and $\mathcal{V}_{M}^{+}(z)$ are defined identically to $\mathcal{V}_{M,n}^{-}(z)$ and $\mathcal{V}_{M,n}^{+}(z)$ after replacing $\widehat\Sigma_{T,n}$, $\widehat\Sigma_{DT,n}$ and $\widehat a_{M,n}$ with $\Sigma_T^*$, $\Sigma_{DT}^*$ and $a_{M}^*$ and $Z_{M}^*$ is defined identically to $Z_{M,n}^*$ after replacing $\widehat\Sigma_{T,n}$, $\widehat\Sigma_{DT,n}$, $D_n(M)$ and $T_n^*(M)$ with $\Sigma_T^*$, $\Sigma_{DT}^*$, $D^*(M)+\mu_D^*(M)$ and $T^*(M)$.  This convergence is joint with that of $(T_{n_s}^*,\widehat{\Sigma}_{n_s},K_{n_s,\gamma},\widehat M_{n_s})$ so that we may write 
\begin{equation}
(T_{n_s}^*,\widehat{\Sigma}_{n_s},K_{n_s,\gamma},\widehat M_{n_s},\mathcal{V}_{M,n_s}^{-}(Z_{M,n_s}^*),\mathcal{V}_{M,n_s}^{+}(Z_{M,n_s}^*))\overset{d}\longrightarrow (T^*,{\Sigma}^*,K_{\gamma}^*,\widehat M,\mathcal{V}_{M}^{-}(Z_{M}^*),\mathcal{V}_{M}^{+}(Z_{M}^*)) \label{eq: joint conv}
\end{equation}
under $\{\mathbb{P}_{n_s}\}$ for any $M\in \mathcal{M}$.

By Lemma 9 of \cite{AKM20}, $F_{TN}(t;\mu,\Sigma_T(M,M),\mathcal{L},\mathcal{U})$ is continuous over the set
\[
\left\{ \left(t,\mu,\Sigma_T(M,M)\right)\in\mathbb{R}^{3},\mathcal{L}\in\mathbb{R}\cup\left\{ -\infty\right\} ,\mathcal{U}\in\mathbb{R}\cup\left\{ \infty\right\} :\Sigma_T(M,M)>0,\mathcal{L}<t<\mathcal{U}\right\}
\]
so that with Assumption 4, \eqref{eq: joint conv} implies
\begin{align}
&\left(F_{TN}(T_{n_s}^*(\widehat{M}_{n_s});0,\widehat\Sigma_{T,n_s}(\widehat{M}_{n_s},\widehat{M}_{n_s}),\max\left\{\mathcal{V}_{\widehat M_{n_s},n_s}^{-}(Z_{\widehat M_{n_s},n_s}^*),-\sqrt{\widehat\Sigma_{T,n_s}(\widehat M_{n_s},\widehat M_{n_s})}K_{n_s,\gamma}\right\},\right. \notag \\
&\qquad\left.\min\left\{\mathcal{V}_{\widehat M_{n_s},n_s}^{+}(Z_{\widehat M_{n_s},n_s}^*),\sqrt{\widehat\Sigma_{T,n_s}(\widehat M_{n_s},\widehat M_{n_s})}K_{n_s,\gamma}\right\},\mathbf{1}(\widehat{M}_{n_s}=M,\mu_{T,n_s}(\widehat{M}_{n_s})\in CI_{n_s,\widehat{M}_{n_s}}^{P,\gamma})\right) \notag \\
&\overset{d}\longrightarrow \left(F_{TN}(T^*(\widehat{M});0,\Sigma_T^*(\widehat{M},\widehat{M}),\max\left\{\mathcal{V}_{\widehat M}^{-}(Z_{\widehat M}^*),-\sqrt{\Sigma_T^*( \widehat M, \widehat M)}K_{\gamma}^*\right\}, \right. \label{eq: trunc normal conv}\\
&\qquad\left.\min\left\{\mathcal{V}_{\widehat M}^{+}(Z_{\widehat M}^*),\sqrt{\Sigma_T^*(\widehat M,\widehat M)}K_{\gamma}^*\right\},\mathbf{1}\left(\widehat{M}=M,-\sqrt{\Sigma_T^*(\widehat M,\widehat M)}K_{\gamma}^*\leq T^*(\widehat M)\leq \sqrt{\Sigma_T^*(\widehat M,\widehat M)}K_{\gamma}^*\right)\right), \notag 
\end{align}
since $\mu_{T,n}(\widehat{M}_n)\in CI_{n,\widehat{M}_n}^{P,\gamma}$ is equivalent to 
\[-\sqrt{\widehat\Sigma_{T,n}(\widehat M_n,\widehat M_n)}K_{n,\gamma}\leq T_n^*(\widehat M_n)\leq \sqrt{\widehat\Sigma_{T,n}(\widehat M_n,\widehat M_n)}K_{n,\gamma}.\]  
Given the equivalence in \eqref{eq: trunc normal equiv}, Lemma 1 and \eqref{eq: trunc normal conv}, the result of the proposition follows from the same arguments used to prove the first part of Corollary 2 of \cite{AKM20}. $\blacksquare$

\textbf{Proof of Propsition 2:}~To see why the first inequality holds, note the following:
\begin{align*}
&\liminf_{n\rightarrow\infty}\inf_{\mathbb{P}\in\mathcal{P}_n}\mathbb{P}\left(\mu_{T,n}(\widehat{M}_n;\mathbb{P})\in CI_{n,\widehat{M}_n}^{H,\alpha}\right) \\
&\geq \liminf_{n\rightarrow\infty}\inf_{\mathbb{P}\in\mathcal{P}_n}\mathbb{P}\left(\mu_{T,n}(\widehat{M}_n;\mathbb{P})\in CI_{n,\widehat{M}_n}^{H,\alpha}\left|\mu_{T,n}(\widehat{M}_n;\mathbb{P})\in CI_{n,\widehat{M}_n}^{P,\gamma}\right.\right)\mathbb{P}\left(\mu_{T,n}(\widehat{M}_n;\mathbb{P})\in CI_{n,\widehat{M}_n}^{P,\gamma}\right) \\
&=\liminf_{n\rightarrow\infty}\inf_{\mathbb{P}\in\mathcal{P}_n}\sum_{M\in\mathcal{M}}\left\{\mathbb{P}\left(\mu_{T,n}(\widehat{M}_n;\mathbb{P})\in CI_{n,\widehat{M}_n}^{H,\alpha}\left|\widehat M_n=M,\mu_{T,n}(\widehat{M}_n;\mathbb{P})\in CI_{n,\widehat{M}_n}^{P,\gamma}\right.\right)\right. \\
&\qquad\qquad\qquad\qquad\quad\times\left.\mathbb{P}\left(\widehat M_n=M,\mu_{T,n}(\widehat{M}_n;\mathbb{P})\in CI_{n,\widehat{M}_n}^{P,\gamma}\right)\right\} \\
&\geq\frac{1-\alpha}{1-\gamma}\liminf_{n\rightarrow\infty}\inf_{\mathbb{P}\in\mathcal{P}_n}\sum_{M\in\mathcal{M}}\mathbb{P}\left(\widehat M_n=M,\mu_{T,n}(\widehat{M}_n;\mathbb{P})\in CI_{n,\widehat{M}_n}^{P,\gamma}\right) \\
&=\frac{1-\alpha}{1-\gamma}\liminf_{n\rightarrow\infty}\inf_{\mathbb{P}\in\mathcal{P}_n}\mathbb{P}\left(\mu_{T,n}(\widehat{M}_n;\mathbb{P})\in CI_{n,\widehat{M}_n}^{P,\gamma}\right)\geq \frac{1-\alpha}{1-\gamma}(1-\gamma)=1-\alpha,
\end{align*}
where the second inequality follows from Lemma 6 of \cite{AKM20} and Proposition 1 and the final inequality holds by Assumption 4.  The second inequality in the proposition follows from essentially the same argument used to prove the final part of Corollary 2 of \cite{AKM20}.   $\blacksquare$ 

\bibliographystyle{apalike}
\bibliography{hybrid_post_sel}